\documentclass[eqsecnum,twocolumn,pre,showpacs]{revtex4}

\usepackage{dcolumn}
\usepackage{amsfonts}
\usepackage{amsmath}
\usepackage{amssymb}
\usepackage{epsfig}
\usepackage{bm}

\begin{document}

\newcommand{\brm}[1]{\bm{{\rm #1}}}
\newcommand{\tens}[1]{\underline{\underline{#1}}}
\newcommand{\mm}{\overset{\leftrightarrow}{m}}
\newcommand{\xv}{\bm{{\rm x}}}
\newcommand{\Rv}{\bm{{\rm R}}}
\newcommand{\uv}{\bm{{\rm u}}}
\newcommand{\nv}{\bm{{\rm n}}}
\newcommand{\Nv}{\bm{{\rm N}}}
\newcommand{\ev}{\bm{{\rm e}}}
\newcommand{\av}{\bm{{\rm a}}}
\newcommand{\bv}{\bm{{\rm b}}}
\newcommand{\tv}{\bm{{\rm t}}}
\newcommand{\Tv}{\bm{{\rm T}}}

\title{Soft elasticity in biaxial smectic and smectic-$C$ elastomers}

\author{Olaf Stenull}
\affiliation{Fachbereich Physik, Universit\"at Duisburg-Essen,
Campus Essen, 45117 Essen, Germany }

\author{T. C. Lubensky}
\affiliation{Department of Physics and Astronomy, University of
Pennsylvania, Philadelphia, PA 19104, USA }

\vspace{10mm}
\date{\today}

\begin{abstract}
\noindent Ideal (monodomain) smectic-$A$ elastomers crosslinked in
the smectic-$A$ phase are simply uniaxial rubbers, provided
deformations are small. From these materials smectic-$C$
elastomers are produced by a cooling through the smectic-$A$ to
smectic-$C$ phase transition. At least in principle, biaxial
smectic elastomers could also be produced via cooling from the
smectic-$A$ to a biaxial smectic phase. These phase transitions,
respectively from $D_{\infty h}$ to $C_{2h}$ and from $D_{\infty
h}$ to $D_{2h}$ symmetry, spontaneously break the rotational
symmetry in the smectic planes. We study the above transitions and
the elasticity of the smectic-$C$ and biaxial phases in three
different but related models: Landau-like phenomenological models
as functions of the Cauchy--Saint-Laurent strain tensor for both
the biaxial and the smectic-$C$ phases  and a detailed model,
including contributions from the elastic network, smectic layer
compression, and smectic-$C$ tilt for the smectic-$C$ phase as a
function of both strain and the $c$-director. We show that the
emergent phases exhibit soft elasticity characterized by the
vanishing of certain elastic moduli. We analyze in some detail the
role of spontaneous symmetry breaking as the origin of soft
elasticity and we discuss different manifestations of softness
like the absence of restoring forces under certain shears and
extensional strains.
\end{abstract}


\maketitle

\section{Introduction}
\label{introduction} Liquid crystalline
elastomers~\cite{WarnerTer2003} are fascinating hybrid materials
that combine the elastic properties of rubber with the
orientational and positional order of liquid
crystals~\cite{deGennesProst93_Chandrasekhar92}. As in
conventional liquid crystals, there exists a great variety of
phases in liquid crystalline elastomers. For example, nematic,
cholesteric, smectic-$A$ (Sm$A$), chiral smectic-$A^*$
(Sm$A^\ast$), smectic-$C$ (Sm$C$), and chiral smectic-$C^\ast$
(Sm$C^\ast$) phases have been created in elastomeric
forms~\cite{WarnerTer2003}. Among these elastomers, nematics have
to date received the most attention leading to the discovery of a
number of remarkable properties of these materials such as soft
elasticity~\cite{golubovic_lubensky_89,FinKun97,VerWar96,War99,LubenskyXin2002},
dynamic soft
elasticity~\cite{terentjev&Co_NEhydrodyn,stenull_lubensky_2004,stenull_lubensky_comment},
and anomalous
elasticity~\cite{stenull_lubensky_epl2003,Xing_Radz_03,stenull_lubensky_anomalousNE_2003}.
In contrast, the understanding of smectic elastomers is much less
developed, at least from a theoretical point of view. Given that
there is a substantial literature treating the synthesis and
experimental properties of smectic elastomers
\cite{shibaev_81_82,fischer_95,bremer&Co_1993,benne&Co_1994,hiraoka&CO_2001,hiraoka&CO_2005,GebhardZen1998,ZentelBre2000,StannariusZen2002},
that smectic elastomers have intriguing properties and potential
for device applications such as manometer-scale
actuators~\cite{lehmann&Co_01}, and that smectics play a leading
role in conventional liquid crystals where they have attracted
outstanding scientific and technological interest since the
discovery of spontaneous ferroelectricity in $C^\ast$
smectics~\cite{meyer&Co_75}, it is clear that a deeper theoretical
understanding of smectic elastomers is desirable.

To date there have been, as far as we know, only relatively few
theoretical investigations of smectic elastomers for the apparent
reason that such investigations are difficult due to the
complexity and low symmetry of the material. Terentjev and Warner
developed expressions for the elastic energies of Sm$A$ and
Sm$A^\ast$ elastomers~\cite{terentjev_warner_SmA_1994} as well as
for Sm$C$ and Sm$C^\ast$
elastomers~\cite{terentjev_warner_SmC_1994} based on group
theoretical arguments. The coupling of the smectic layers to the
elastic network was critically discussed and shortcomings of
Ref.~\cite{terentjev_warner_SmA_1994} in this respect were
corrected in~\cite{lubensky&Co_94}. Subsequently, the damping
effect of the rubber-elastic matrix on the fluctuations of the
smectic layers, leading to a suppression of the Landau-Peierls
instability and true one-dimensional long-range order, was
analyzed in~\cite{terentjev_warner_lubensky_95}. Weilepp and
Brand~\cite{weilepp_brand_1998} discussed an undulation
instability as a possible explanation for the turbidity of Sm$A$
elastomers under stretch along the normal of the smectic layers.
Osborne and Terentjev~\cite{osborne_terentjev_2000} derived
expressions for the effective elastic constants of Sm$A$
elastomers when these are viewed as effectively uniaxial systems
and revisited the suppression of fluctuations of the smectic
layers. Very recently, Adams and Warner
(AW)~\cite{adams_warner_2005,adams_warner_SmC_2005,adams_warner_2006}
set up a model for the elasticity of smectic elastomers by
extending the so-called neoclassic model of rubber elasticity,
which was originally developed and very successfully used to
describe nematic elastomers~\cite{WarnerTer2003}, to include the
effects of smectic layering. Also just recently we worked out
theories for the low-frequency long-wavelength dynamics of smectic
elastomers in respectively the Sm$A$, biaxial and Sm$C$
phases~\cite{stenull_lubensky_SmCdynamics}.

In principle, smectic elastomers can be produced either by cooling
nematic elastomers through the transition to the smectic phase or
by crosslinking in the smectic phase. Both Sm$A$ and Sm$C$
elastomers have been prepared mostly by the second method, e.g.,
by crosslinking side chain liquid crystalline polymers, which have
a tendency to form layers because their mesogens are often
immiscible with the polymeric backbone~\cite{shibaev_81_82}, or by
crosslinking polymer chains or hydrophobic tails in bilayer
lamellar phases of respectively diblock copolymers or surfactant
molecules~\cite{fischer_95}. Crosslinking in the smectic phase
tends to lock the smectic layers to the crosslinked
network~\cite{lubensky&Co_94}. Without this lock-in, the phase of
the smectic mass-density-wave can translate freely relative to the
elastomer as it can in smectics in
aerogels~\cite{radzihovsky_toner_99}. To keep our discussion as
simple as possible, we will not consider in the following the case
of crosslinking in the nematic phase, and we take the lock-in of
the smectic layers and the elastic matrix as given.

As in nematics, unconventional properties are most pronounced in
samples of smectic elastomers that have an ideal, monodomain
morphology so that the director has a uniform orientation
throughout. In practice, however, liquid crystal elastomers tend
to be non-ideal, i.e.\ the material segregates into many domains,
each having its own local director. In order to avoid such
polydomain samples, elaborate crosslinking schemes involving
electric or mechanical external fields for aligning the director
have been developed and successfully applied to smectic
elastomers~\cite{bremer&Co_1993,benne&Co_1994,hiraoka&CO_2001}.
The latest achievement in this respect was reported very recently
by Hiraoka {\em et al}.~\cite{hiraoka&CO_2005}, who produced a
monodomain sample of a Sm$C$ elastomer forming spontaneously from
a Sm$A$ phase upon cooling and carried out experiments
demonstrating its spontaneous and reversible deformation in a
heating an cooling process.

A monodomain Sm$A$ elastomer crosslinked in the Sm$A$ phase is
effectively a uniaxial solid with $D_{\infty h}$ symmetry, at
least for small deformations. For lager deformations, however,
Sm$A$ elastomers can show unconventional effects, as do Sm$A^\ast$
elastomers in external electric fields. These unconventional
effects in Sm$A$ elastomers are outside the scope of this paper
and will be addressed in a separate
paper~\cite{stenull_lubensky_smA_2006}.

\begin{figure}
\centerline{\includegraphics[width=8.4cm]{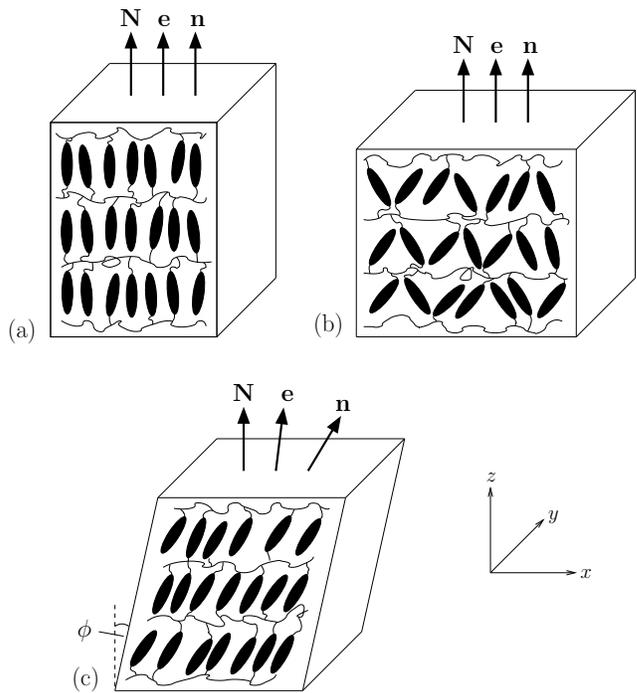}}
\caption{Sample distortion and rotations of the Frank director
$\nv$, the uniaxial anisotropy axis $\ev \equiv \ev_z$, and the
layer normal $\Nv$ in a transition from a (a) Sm$A$ to (b) a
biaxial and (c) a sheared Sm$C$ elastomer. In part (c) we chose
the geometry, i.e., the coordinate system in target space, so that
the smectic layers do not rotate.} \label{fig1}
\end{figure}

The elastic properties of a Sm$C$ elastomer depend on whether it
was crosslinked in the Sm$A$ or Sm$C$ phase. If it is prepared by
crosslinking in a Sm$C$ phase, reached either by stretching or by
applying an external electric field, it is a conventional biaxial
solid with $C_{2h}$ symmetry. If, however, the Sm$C$ phase
develops spontaneously upon cooling from a uniaxial Sm$A$ (as in
the work of Hiraoka {\em et al}.) then the underlying phase
transition from $D_{\infty h}$ to $C_{2h}$ symmetry (see
Fig.~\ref{fig1}) spontaneously breaks the continuous rotational
symmetry in the smectic planes. As a consequence of the Goldstone
theorem that requires any phase with a spontaneously broken
continuous symmetry to have modes whose energy vanishes with
wavenumber, like monodomain nematic
elastomers~\cite{golubovic_lubensky_89,FinKun97,VerWar96,War99,LubenskyXin2002},
these Sm$C$ elastomers are predicted to exhibit soft elasticity
characterized by the vanishing of a certain elastic modulus and
the associated absence of restoring forces to strains along
specific symmetry directions~\cite{stenull_lubensky_letter_2005}.

Though biaxial phases are notoriously hard to find in nature, it
is possible, at least in principle, that a biaxial Sm$A$ phase
spontaneously forms upon cooling from a Sm$A$ elastomer. In what follows,
we will often simply refer to biaxial Sm$A$ elastomers as biaxial smectic elastomers or biaxial elastomers. In contrast to the aforementioned phase transition to a soft Sm$C$
elastomer, the transition to the biaxial Sm$A$ phase involves no net tilt of the mesogens
and takes the system to $D_{2h}$ instead of $C_{2h}$ symmetry (see
Fig.~\ref{fig1}). Nontheless, this transition also breaks the
rotational invariance in the smectic layers spontaneously and thus
the emerging biaxial phase is
soft~\cite{stenull_lubensky_letter_2005} similar to ideal nematic
and Sm$C$ elastomers.

In this paper we study the phase transitions from Sm$A$ elastomers
to biaxial and Sm$C$ elastomers and the elasticity of the emergent
phases. As briefly presented in
Ref.~\cite{stenull_lubensky_letter_2005}, we set up three
different but related models. Our first two models involve only
elastic degrees of freedom, i.e., they involve exclusively the
usual strain tensor. Our third model also includes the Frank
director specifying the direction of local molecular order. Each
of the models is analyzed within mean-field theory, revealing as
the primary result the soft elasticity of biaxial and Sm$C$
elastomers
\cite{stenull_lubensky_letter_2005,adams_warner_SmC_2005,adams_warner_2006}
alluded to in the forgoing paragraphs.

The outline of our paper is as follows:
Section~\ref{lagrangeElasticity} briefly reviews the Lagrange
formulation of elasticity theory in the context of uniaxial
elastomers to establish notation and to provide a starting point
for our models to follow. Section~\ref{biaxialElastomers} presents
our strain-only theory for biaxial elastomers. We study the
transition from the Sm$A$ to the biaxial state and the elastic
properties of the latter. We derive the elastic energy density of
the biaxial state and discuss its softness with respect to certain
shears and extensional strains. Section~\ref{smecticElastomers1}
contains our strain-only theory for Sm$C$ elastomers. We
investigate the Sm$A$-to-Sm$C$ transition and calculate the
elastic energy density of the Sm$C$ phase. Different
manifestations of the softness of Sm$C$ elastomers are pointed
out. Section~\ref{smecticElastomers2} formulates our theory for
Sm$C$ elastomers with strain, Frank director, and smectic layers.
Using the polar decomposition theorem, we derive transformations
between vectors that transform according to operations on
reference-space positions $\xv$ of the undistorted medium and
those that transform according to operations on target-space
positions $\Rv(\xv)$ of the distorted medium, and we formulate the
elastic energy for coupled director and strain in terms of
nonlinear-strain and director fields that transform under
reference-state operations only.  In this approach, phase
transitions can be studied without specifying actual orientation
in space. We develop a full model free energy that includes
contributions from the crosslinked network, smectic layer
compression, and coupling between the Frank director and the
smectic layer normal. Then, as above, we study the phase
transition from the Sm$A$ to the Sm$C$ phase and the elastic
energy density of the emergent phase. In addition, we discuss the
general form of soft deformations and strains in Sm$C$ elastomers
based on rotational invariance in the smectic planes and we
elaborate on softness under extensional strains. Concluding
remarks are given in Sec.~\ref{concludingRemarks}. There are in
total 4 appendixes which contain technical details or arguments
that lie somewhat aside the line of thought of the main text.

\section{Lagrangian description of uniaxial elastomers}
\label{lagrangeElasticity} As argued above, ideal Sm$A$ elastomers
are  macroscopically simply uniaxial rubbers, but with nonlinear
properties that distinguish them from simple uniaxial solids. We
will employ the usual Langrangian
formalism~\cite{Landau-elas,tomsBook} of elasticity theory. Here,
we briefly review key elements of this formalism in the context of
uniaxial elastomers to establish notation and to provide some
background information.

In the Langrangian formalism mass points in an undistorted medium
(or body), which we take as the reference space, are labelled by
vectors $\xv$. Mass points of the distorted medium are at
positions
\begin{equation}
\Rv(\xv) = \xv + \uv(\xv) ,
\end{equation}
that constitute what we call the target space.  Both reference
space points $\xv$ and target space points $\Rv(\xv)$ exist in the
same physical Euclidean space $\cal E$ where measurements are
made. Thus, $\Rv(\xv)$ is a mapping from $\cal E$ to $\cal E$.
Both $\xv$ and $\Rv(\xv)$ can be decomposed into components along
the standard orthonormal basis $\{\av_i|i = x,y,z\}$ of $\cal E$:
\begin{equation}
\xv = x_i \, \av_i, \qquad \Rv(\xv) = R_i ( \xv ) \, \av_i .
\end{equation}
Here and in what follows, we use the summation convention on
repeated indices unless we indicate otherwise.  We will also use
the convention that indices from the middle of the alphabet run
over all space coordinates, $i,j,k = x,y,z$. We choose our
coordinate system so that the $z$-axis is along the uniaxial
direction of the initial reference material. Indices from the
beginning of the alphabet, $a$, $b$, $c$, run over $x$ and $y$
only, i.e., over directions perpendicular to the anisotropy axis.

Though reference- and target-space vectors both exist in $\cal E$,
they transform under distinct and independent transformation
operations. Let $\tens{O}_R$ and $\tens{O}_T$ denote,
respectively, matrices describing transformations (which we will
take mostly to be rotations but which could include reflections
and inversions as well) in the reference and target spaces; then
under these transformations, $\xv \rightarrow \xv' = \tens{O}_R
\xv$ and $\Rv(\xv) \rightarrow \Rv'(\xv) = \tens{O}_T \Rv(\xv)$,
or in terms of components relative to the $\av$-basis
\begin{subequations}
\begin{eqnarray}
R'_i( \xv ) & = &O_{T, ij} R_j ( \xv )\label{eq:transform1} \, ,
 \\
x'_i & = & O_{R,ij} \, x_j \,  . \label{eq:transform2}
\end{eqnarray}
\label{eq:transform}
\end{subequations}
Unless otherwise specified, we will view $\tens{O}_R$ and
$\tens{O}_T$ as operators that rotate vectors rather than
coordinate systems.

Elastic energies are invariant under arbitrary rigid rotations and
translations in the target space and under symmetry operations of
the references space of the form $\xv\rightarrow \xv' =
\tens{O}_R^{-1} \xv+ \bv$, where $\bv$ is a constant vector and
$\tens{O}_R$  is a matrix associated with some symmetry element of
the reference space.  Thus, elastic energies are invariant under
transformations of the form
\begin{equation}
\Rv(\xv) \rightarrow \Rv'(\xv') = \tens{O}_T \Rv(\tens{O}_R^{-1}
\xv + \bv ) + {\bm{{\rm X}}} , \label{eq:transformR}
\end{equation}
where $\tens{O}_T$ is an arbitrary target-space rotation matrix
and ${\bm{{\rm X}}}$ is a constant displacement vector.  In what
follows, we will generally ignore the displacements ${\bm{{\rm
X}}}$ and $\bv$. The use of $\tens{O}_R^{-1}$ in
Eq.~(\ref{eq:transformR}) rather than $\tens{O}_R$ is a matter of
convention~\cite{footnoteRotationConnvention}.  With the choice
$\tens{O}_R^{-1}$, when $\tens{O}_T = \tens{\delta}$ where
$\tens{\delta}$ is the unit matrix, the mapping $\Rv'(\xv)$ takes
the point $\tens{O}_R \xv$ to the same point in $\cal E$ as the
mapping $\Rv(\xv)$ takes the point $\xv$.

We will usually represent the the reference-space points $\xv$ in
terms of their coordinates relative to the basis $\{\av_i\}$.  We
will, however, find it useful on occasion to consider orthonormal
bases locked to the reference medium and to represent
reference-space vectors relative to them.  The initial basis
$\{\tilde{\ev}_i|i = x,y,z\}$ is identical to the basis $\{\av_i|i
= x,y,z\}$, and $\xv= x_i \av_i \equiv x_i \tilde{\ev}_i$.
$\tilde{\ev}_z$ is thus a vector along the uniaxial axis of the
undistorted body \cite{e-n0}.  Under rotations of the body basis,
$\xv= x'_i \tilde{\ev}'_i = x_i \ev_i$, where
\begin{equation}
\tilde{\ev}'_i = O_{R,ij} \tilde{\ev}_j \label{eq:basischange}
\end{equation}
and $x'_i = O_{R,ij} x_j$. As we shall discuss in more detail in
Sec.~\ref{sec:polarDecomposition} associated with each
reference-space vector, there is a target-space vector of the same
length. Thus, associated with the reference basis
$\{\tilde{\ev}_i\}$, there is a target basis $\{\ev_i\}$. In
particular there is a target-space anisotropy direction $\ev_z$
associated with $\tilde{\ev}_z$.

Distortions of the reference medium are described by the Cauchy
deformation tensor $\tens{\Lambda}$ with components
\begin{align}
\Lambda_{ij} =
\partial R_i /\partial x_j \equiv
\partial_j R_i \, .
\end{align}
It transforms under the operations of Eq.~(\ref{eq:transform})
according to
\begin{equation}
\tens{\Lambda} ( \xv ) \rightarrow \tens{O}_T \,
\tens{\Lambda}(\xv^\prime ) \, \tens{O}_R^{-1} \,
\end{equation}
i.e., the right subscript transforms under target-space rules and
the left under reference-space rules. Usually, Lagrangian elastic
energies are expressed in terms of the Cauchy-Saint-Venant
\cite{Love1944,Landau-elas} nonlinear strain tensor $\tens{u}=
(\tens{g} - \tens{\delta})/2$, where
\begin{equation}
\tens{g} = \tens{\Lambda}^T \tens{\Lambda}
\end{equation}
is the metric tensor. The components of $\tens{u}$ are
\begin{subequations}
\label{defStrain}
\begin{align}
\label{defStraina} u_{ij} ( \xv ) &= \textstyle{\frac{1}{2}}\,
(\Lambda^T_{ik}\Lambda_{kj} - \delta_{ij})
 \\
 \label{defStrainb}
&=  \textstyle{\frac{1}{2}} \, (\partial_i u_j +\partial_j u_i +
\partial_i u_k\partial_j u_k)\, .
\end{align}
\end{subequations}
The strain $\tens{u}$ is a reference-space tensor: it is invariant
under transformations $\tens{O}_T$ in the target space, but it
transforms like a tensor under reference-space transformations:
\begin{align}
\label{baseRotU} \tens{u}(\xv) \rightarrow \tens{O}_R\,
\tens{u}(\xv^\prime) \, \tens{O}_R^{-1} \, .
\end{align}
This expression applies both to physical transformations of
reference-space vectors under $x'_i = O_{R,ij}^{-1} x_j$ or under
changes of basis described by Eq.~(\ref{eq:basischange}) under
which $\Lambda_{ij} \rightarrow \Lambda'_{ij} = \partial R_i/
\partial x'_j = \Lambda_{ik} O_{R,kj}^{-1}$.

To discuss incompressible materials, such as most elastomers, it
can be more appropriate to use variables other than the strain
tensor to account for deformations that are not pure shear. In the
case of uniaxial elastomers, such variables are the relative
change of the system volume $V$,
\begin{subequations}
\begin{align}
\label{defEta} \eta \equiv \delta V /V = [\det \tens{\Lambda}^T
\tens{\Lambda}]^{1/2} -1 = [(\det(1 + 2 \tens{u})]^{1/2} -1\, ,
\end{align}
and the relative change of separation of mass points whose
separation vector in the reference state is along the $z$ axis:
\begin{align}
\label{defEtaZ} \eta_z \equiv \delta L_z /L_z =  |\Lambda_{iz}
\Lambda_{iz}|^{1/2} - 1= (1 + 2 u_{zz})^{1/2} - 1\, .
\end{align}
\end{subequations}
Using these variables and, where appropriate, the elements of the
strain tensor, the elastic free energy density of a uniaxial
elastomer to harmonic order can be expressed as
\begin{align}
\label{uniEnNonlinear} f_{\text{uni}} &= \textstyle{\frac{1}{2}}
\, C_1\, \eta_{z}^2 + C_2 \, \eta_{z} \eta +
\textstyle{\frac{1}{2}} \, C_3\, \eta^2
\nonumber \\
&+ C_4 \, \hat{u}_{ab}^2 + C_5 \,u_{az}^2,
\end{align}
where
\begin{align}
\label{defUhat} \hat{u}_{ab}=u_{ab}-\ \textstyle{\frac{1}{2}}\,
\delta_{ab} u_{cc}
\end{align}
is the two-dimensional symmetric, traceless strain tensor with
two-independent components that can be expressed as
$\tens{\hat{u}}=u_1 (\tilde{\ev}_x \tilde{\ev}_x-\tilde{\ev}_y
\tilde{\ev}_y) + u_2 (\tilde{\ev}_x \tilde{\ev}_y + \tilde{\ev}_y
\tilde{\ev}_x)$. The elastic constant $C_1$ describes dilation or
compression along $z$ and $C_3$ describes expansion or compression
of the bulk volume. $C_2$ couples these two types of deformations.
$C_4$ and $C_5$ respectively describe shears in the plane
perpendicular to the anisotropy axis and in the planes containing
it. With the variables used in Eq.~(\ref{uniEnNonlinear}) it is
evident that the incompressible limit corresponds to $C_3
\rightarrow \infty$.

If one approximates $\eta$ and $\eta_z$ by the respective leading
terms in the strains one is left with
\begin{subequations}
\label{linearApprox}
\begin{align}
\eta &= u_{ii} + O (  u_{ij}^2 )\, ,
\\
\eta_z &= u_{zz} + O (  u_{zz}^2 ) \, .
\end{align}
\end{subequations}
and the elastic energy density~(\ref{uniEnNonlinear}) reduces to
the more standard-type expression
\begin{align}
\label{uniEn} f_{\text{uni}} &= \textstyle{\frac{1}{2}} \, C_1\,
u_{zz}^2 + C_2 \, u_{zz} u_{ii} +  \textstyle{\frac{1}{2}} \,
C_3\, u_{ii}^2
\nonumber \\
&+ C_4 \, \hat{u}_{ab}^2 + C_5 \,u_{az}^2.
\end{align}
Our models to be presented in the following are in spirit Landau
expansions in powers of $u_{ij}$ or in powers of  $u_{ij}$ and the
Frank director, respectively. Hence, for our purposes, the
approximations in Eqs.~(\ref{linearApprox}) will be sufficient,
and we can use Eq.~(\ref{uniEn}) as a starting point for the
construction of our models. As we show in
Appenix~\ref{app:nonlinear}, using Eq.~(\ref{uniEn}), instead of
the more general elastic energy density
Eq.~(\ref{uniEnNonlinear}), leaves our results qualitatively
unchanged, though the more general theory is needed for a correct
description of the incompressible limit.

\section{Biaxial smectic-$A$ elastomers -- strain-only theory}
\label{biaxialElastomers}

In our first theory we consider the case that the shear modulus
$C_4$ becomes negative as it will in response to biaxial ordering
of the constituent mesogens of a uniaxial Sm$A$ elastomer.

\subsection{Phase transition from uniaxial to biaxial elastomers}
\label{biaxialPhaseTransition} If $C_4$ becomes negative, order of
the shear strain $\hat{u}_{ab}$ sets in, and higher-order terms
featuring $\hat{u}_{ab}$ have to be added to Eq.~(\ref{uniEn})
which leads to the model elastic energy density
\begin{align}
\label{uniEn1} f_{\text{uni}}^{(1)} = f_{\text{uni}} + A_1  \,
u_{zz} \hat{u}_{ab}^2 + A_2 \,  u_{ii} \hat{u}_{ab}^2 + B \,
(\hat{u}_{ab}^2)^2 ,
\end{align}
where we have dropped qualitatively inconsequential higher order
terms. For the analysis that follows it is useful to regroup the
terms in $f_{\text{uni}}^{(1)}$ by completing the squares in
$\frac{1}{2} C_1 u_{zz}^2 + A_1 u_{zz} \hat{u}_{ab}^2$, etc., and
to reexpress it as a sum of two terms,
\begin{align}
f_{\text{uni}}^{(1)} = f_{\text{uni}}^{(1,1)} +
f_{\text{uni}}^{(1,2)} ,
\end{align}
where
\begin{subequations}
\begin{align}
\label{f^11} f_{\text{uni}}^{(1,1)}& =  \textstyle{\frac{1}{2}} \,
C_1 v_{zz}^2 + C_2\,  v_{zz} v_{ii} +\textstyle{\frac{1}{2}} \,
C_3 \, v_{ii}^2 + C_5 \, u_{az}^2
\\
\label{f^12} f_{\text{uni}}^{(1,2)}& =  C_4\, \hat{u}_{ab}^2 + B_R
\,
(\hat{u}_{ab}^2)^2 \\
& = 2C_4 (u_1^2 + u_2^2) +4 B_R (u_1^2 + u_2^2)^2.
\end{align}
\end{subequations}
The energy $f_{\text{uni}}^{(1,2)}$ is clearly identical to the
energy of an $xy$ model with a two component vector $(u_1,u_2)$.
Here, we have introduced the composite strains
\begin{subequations}
\label{compStriansV}
\begin{align}
v_{zz}&= u_{zz}- \alpha \, \hat{u}_{ab}^2 \, ,
\\
v_{ii}& = u_{ii} - \beta \, \hat{u}_{ab}^2 \, ,
\end{align}
\end{subequations}
where $\alpha$ and $\beta$ are combinations of the coefficients in
$f_{\text{uni}}^{(1)}$,
\begin{align}
\label{resAlphaBeta} \left(
\begin{array}{c}
\alpha
\\
\beta
\end{array}
\right) = \frac{-1}{C_1\, C_3 - C_2^2} \left(
\begin{array}{c}
C_3 A_1 - C_2 A_2
\\
C_1 A_2 - C_2 A_1
\end{array}
\right) .
\end{align}
The subscript $R$ in Eq.~(\ref{f^12}) indicates that the elastic
constant $B$ is renormalized by the ordering of $\hat{u}_{ab}$:
\begin{align}
B_R = B - \textstyle{\frac{1}{2}} \alpha^2\,  C_1 - \alpha \beta
\, C_2 - \textstyle{\frac{1}{2}} \beta^2 \, C_3 \, .
\end{align}
Note that the coefficient $\beta$ vanishes in the limit $C_3 \to
\infty$. $\alpha$ and $B_R$, on the other hand, remain non-zero.
We will assume that $B_R$ remains positive.  If it did not, we
would have to add higher-order terms in $\hat{u}_{ab}$, and the
transition to the biaxial phase would be first order.

Now we determine the possible equilibrium states $\tens{u}^0$ of
our model by minimizing $f_{\text{uni}}^{(1)}$. As is evident from
Eq.~(\ref{f^11}), $f_{\text{uni}}^{(1)}$ is minimized for a given
equilibrium value $\hat{u}_{ab}^0$ of $\hat{u}_{ab}$ when
\begin{subequations}
\begin{align}
u_{zz}^0 &=  \alpha \, ( \hat{u}_{ab}^0)^2 \, ,
\\
u_{ii}^0  &=  \beta \, ( \hat{u}_{ab}^0)^2 \, ,
\end{align}
as well as
\begin{align}
u_{az}^0 = 0 \, .
\end{align}
\end{subequations}
The equilibrium value of $\hat{u}_{ab}$ minimizes
$f_{\text{uni}}^{(1,2)}$ and is determined by the equation of
state
\begin{align}
\label{EQS} C_4 \, \hat{u}_{ab}^0 + 2\, B_R \, \hat{u}_{ab}^0
(\hat{u}_{cc^\prime}^0)^2  = 0 \, .
\end{align}
This equation of state is solved by a $\hat{u}_{ab}^0$ of the form
\begin{align}
\label{formUhat} \hat{u}_{ab}^0 = S\left(\tilde{c}_a \tilde{c}_b -
\textstyle{\frac{1}{2}} \delta_{ab} \right) \, ,
\end{align}
where $\tilde{\brm{c}}$ is any unit vector in the $xy$-plane and
where $S$ is a scalar order parameter that takes on the values
\begin{align}
S = \left\{
\begin{array}{ccc}
0& \quad \mbox{for} \quad & C_4 >0 \, ,
\\
\pm \sqrt{-C_4 / B_R}& \quad \mbox{for} \quad& C_4 <0 \, .
\end{array}
\right.
\end{align}
For simplicity, we choose our coordinate system so that the
$x$-axis is along $\tilde{\brm{c}}$. Exploiting
definition~(\ref{defUhat}) and Eq.~(\ref{formUhat}) and taking
$\tilde{\brm{c}} =\tilde{\ev}_x$, we find that the equilibrium
strain tensor $\tens{u}^0$ of the new state for $C_4 <0$ is
diagonal with diagonal-elements
\begin{subequations}
\begin{align}
u_{xx}^0 &= \frac{1}{2} S + \frac{1}{4}(\beta -\alpha) S^2  \, ,
\\
u_{yy}^0  &=  - \frac{1}{2} S + \frac{1}{4} (\beta -\alpha)S^2 \,
,
\\
u_{zz}^0 &= \frac{1}{2}\alpha S^2 \, .
\end{align}
\end{subequations}
Thus, the new state is biaxial with $D_{2h}$ symmetry.

The strain $\tens{u}^0$ provides a complete description of the
macroscopic equilibrium state after the phase transition to the
biaxial state, but it provides no information about a sample's
specific orientation in space.  The latter information is
contained in the Cauchy deformation tensor
\begin{align}
\Lambda_{ij}^0 =
\partial R_i^0 /\partial x_j  \, ,
\end{align}
which is related to $\tens{u}^0$ via
\begin{align}
\tens{u}^0= \textstyle{\frac{1}{2}} \,
(\tens{\Lambda}^{0T}\tens{\Lambda}^0 - \tens{\delta}) \, .
\end{align}
Note, that the equilibrium deformation tensor $\tens{\Lambda}^{0}$
is not uniquely determined by $\tens{u}^0$ since rotations in the
target space change $\tens{\Lambda}^0$ but do not change
$\tens{u}^0$. Because  $\tens{u}^0$ is diagonal, it is natural in
the present case not to rotate the strain after the transition.
Then, $\tens{\Lambda}^0$ is also diagonal with diagonal elements
given by
\begin{subequations}
\begin{align}
\Lambda_{xx}^0 &= \sqrt{ 1+  S + \textstyle{\frac{1}{2}}(\beta
-\alpha) S^2}  \, ,
\\
\Lambda_{yy}^0 &=   \sqrt{ 1 -  S + \textstyle{\frac{1}{2}} (\beta
-\alpha)S^2} \, ,
\\
\Lambda_{zz}^0 &=  \sqrt{ 1+  \alpha S^2} \, .
\end{align}
\end{subequations}
It is worth noting that the limit $C_3\rightarrow \infty$ in which
$\beta$ but not $\alpha$ becomes infinite does not yield the
incompressibility condition $\det \tens{\Lambda} = 1$.  This is
because in our model, $C_3$ multiplies $u_{ii}^2$ and not $(\eta
-1)^2$ [See Eq.(\ref{uniEnNonlinear})].

The emergent anisotropy of the new state in the $xy$-plane can be
characterized by the anisotropy ratio
\begin{align}
r_\perp = \left( \frac{\Lambda_{xx}^0}{\Lambda_{yy}^0} \right)^2
.
\end{align}
Having the equilibrium deformation tensor and the anisotropy ratio
we can express the scalar order parameter $S$ as
\begin{align}
S =  \textstyle{\frac{1}{2}} \, [(\Lambda_{xx}^0)^2 -
(\Lambda_{yy}^0)^2] = \textstyle{\frac{1}{2}} \,(\Lambda_{yy}^0)^2
\, (r_\perp -1 ) \, .
\end{align}
In other words, $S$ is a direct measure for the spontaneous
anisotropy in the $xy$-plane.

\subsection{Elasticity of the biaxial phase}
\label{biaxElas} To determine the elastic properties of the new
state, we expand $f_{\text{uni}}^{(1)}$ in powers of
\begin{align}
\delta \tens{u} = \tens{u} - \tens{u}^0 \, .
\end{align}
Since the equilibrium values of $v_{zz}$, $v_{ii}$ and $u_{az}$
are zero, the expansion of $f_{\text{uni}}^{(1,1)}$ is trivial,
\begin{align}
\label{f^11expanded} \delta f_{\text{uni}}^{(1,1)}& =
\textstyle{\frac{1}{2}} \, C_1 (\delta v_{zz})^2 + C_2\,  \delta
v_{zz} \delta v_{ii} +\textstyle{\frac{1}{2}} \, C_3 \, (\delta
v_{ii})^2
\nonumber \\
&+ C_5 \, (\delta u_{az})^2 ,
\end{align}
where, up to linear order in $\delta u_{ij}$,
\begin{subequations}
\begin{align}
& \delta v_{zz} = \delta u_{zz} - \alpha S (\delta u_{xx} - \delta
u_{yy})   \, ,
\\
& \delta v_{ii}  =  \delta u_{zz} + (1-\beta S) \delta u_{xx} +
(1+\beta S) \delta u_{yy} \, ,
\\
& \delta u_{az} = u_{az} \, .
\end{align}
\end{subequations}
As discussed after Eq.~(\ref{f^12}), the structure of
$f_{\text{uni}}^{(1,2)}$ is identical to that of an $xy$ model,
which has no restoring force perpendicular to the direction of
spontaneous order, which we take to be along the $\tilde{\ev}_x$
direction. Thus with order producing a nonvanishing $u_1 = (u_{xx}
- u_{yy})/2$, there is no restoring force for $u_2 = u_{xy}$,  and
\begin{align}
\label{f^12expanded} \delta f_{\text{uni}}^{(1,2)} =  B_R S^2 (
\delta u_{xx} - \delta{u}_{yy})^2 .
\end{align}
Thus $\delta f_{\text{uni}}^{(1)}$ does not depend on $\delta
u_{xy}$ to harmonic order, and we can conclude already at this
stage that the system is soft with respect to shears in the $xy$
plane of the original reference material. Merging
Eqs.~(\ref{f^11expanded}) and (\ref{f^12expanded}) and after
expressing $\alpha$, $\beta$ and $B_R$ in terms of the original
elastic constants, we obtain
\begin{align}
\delta  f^{(1)} &=  \textstyle{\frac{1}{2}} \, [C_1 + 2C_2 + C_3]
\, (\delta u_{zz})^2 + C_5 \, (\delta u_{az})^2
\nonumber \\
&+ [C_2 + C_3 + (A_1 + A_2)S] \, \delta u_{zz} \delta u_{xx}
\nonumber \\
&+  [C_2 + C_3 - (A_1 + A_2)S] \, \delta u_{zz} \delta u_{yy}
 \nonumber \\
&+  \textstyle{\frac{1}{2}} \, [C_3 + 2 A_2 S + 2 B S^2] \,
(\delta u_{xx})^2
\nonumber \\
&+  \textstyle{\frac{1}{2}} \, [C_3 - 2 A_2 S + 2 B S^2] \,
(\delta u_{yy})^2
\nonumber \\
&+  [C_3 - 2 B S^2] \, \delta u_{xx} \delta u_{yy} \,
\label{deltaf1}
\end{align}
after some algebra.

The strain $\delta \tens{u}$ describes distortions relative to the
new biaxial reference state measured in the coordinates of the
original uniaxial state. It is customary and more intuitive,
however, to express the elastic energy in terms of a strain
\begin{align}
\label{defuPrime} \tens{u}'= (\tens{\Lambda}^{0T})^{-1} \delta
\tens{u} \, (\tens{\Lambda}^{0})^{-1}
\end{align}
measured in the coordinates $x'_i = x_i + u_i^0 = \Lambda^0_{ij}
x_j$ of the new biaxial state. In terms of $\tens{u}'$, $\delta
f^{(1)}$ becomes
\begin{align}
\label{biaxEn} f_{D_{2h}}^{\text{soft}}  &=
\textstyle{\frac{1}{2}} \, C_{zzzz}\, (u^\prime_{zz})^2 +
\textstyle{\frac{1}{2}} \, C_{xzxz} \, (u^\prime_{xz})^2 +
\textstyle{\frac{1}{2}} \, C_{yzyz} \, (u^\prime_{yz})^2
\nonumber \\
&+ C_{zzxx} \, u^\prime_{zz} u^\prime_{xx} + C_{zzyy} \,
u^\prime_{zz} u^\prime_{yy} + \textstyle{\frac{1}{2}} \,
C_{xxxx}\, (u^\prime_{xx})^2
\nonumber \\
&+\textstyle{\frac{1}{2}} \, C_{yyyy}\, (u^\prime_{yy})^2 +
C_{xxyy} \, u^\prime_{xx} u^\prime_{yy} \, .
\end{align}
Our results for the elastic constants $C_{ijkl}$ are listed in
appendix~\ref{app:elasticConstantsBiaxial}. These elastic
constants depend on the original elastic constants featured in
Eq.~(\ref{uniEn1}) and the order parameter $S$ and one retrieves
the uniaxial elastic energy density~(\ref{uniEn}) for $S \to 0$.

Equation (\ref{deltaf1}) highlights a problem with approximating
$\eta = \delta V/V$ with its linearized form $u_{ii}$. In
Eq.~(\ref{uniEnNonlinear}), the limit $C_3 \rightarrow \infty$
enforces incompressibility, i.e., no volume change, even if there
is a phase transition.  In Eq.~(\ref{uniEn}), on the other hand,
this limit only keeps $u_{ii} =0$, and $u_{ii}$ does not measure a
volume change relative to a state whose shape has been changed
because of a phase transition. This is easily seen by noting that
$\delta u_{ii} = \Lambda_{zz}^0 u'_{zz} + \Lambda_{xx}^0 u'_{xx} +
\Lambda_{yy}^0 u'_{yy}$ is not proportional to $u'_{ii}$. Thus,
the limit $C_3 \rightarrow \infty$ does not enforce $\delta V/V =
0$ in the biaxial phase. If the full nonlinear theory of Eq.\
(\ref{uniEnNonlinear}) is used, $C_3$ multiplies $(\delta V/V)^2$
in both the uniaxial and biaxial phases as we will show in
App.~\ref{app:nonlinear}.  In what follows, we will continue to
use free energies that are harmonic in non-ordering nonlinear
strains because they give rise to far less algebraic complexity
than do the more complete theories.  The important feature of soft
elasticity and other physical quantities are not sensitive to
which theory we use.  If detailed treatment of incompressibility
is important, the more complete theory can always be used.

\begin{figure}
\centerline{\includegraphics[width=7cm]{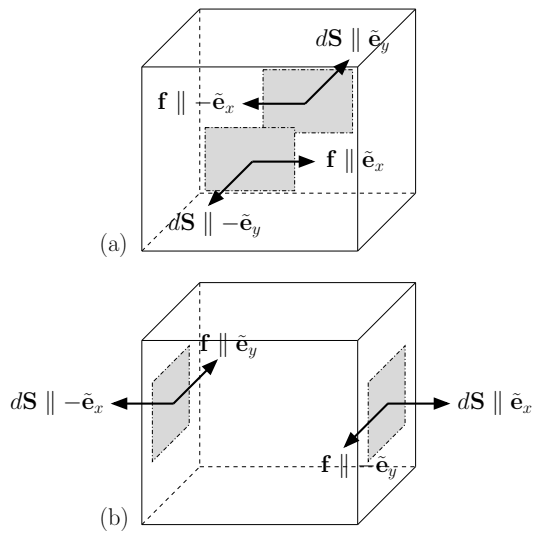}}
\caption{Soft shears in biaxial elastomers. The force $\brm{f}$
exerted across an infinitesimal surface element $d\brm{S}$ is $f_i
= \sigma_{ij} dS_j$. There are no restoring forces for (a)
external forces along $\pm \tilde{\ev}_x$ applied to opposing
surfaces with normal along $\pm \tilde{\ev}_y$, and (b) for
external forces  along $\pm \tilde{\ev}_y$ applied to surfaces
with normal along $\pm \tilde{\ev}_x$.}
\label{fig:biaxialSoftness}
\end{figure}

Because there was no $\delta u_{xy}$ term in the expansion of
$f_{\text{uni}}^{(1)}$, there is no term
\begin{align}
\label{conventionalTerm}
 C_{xyxy} \, (u_{xy}^\prime)^2
\end{align}
as there would be in conventional orthorhombic systems, because
the elastic constant $C_{xyxy}$ is zero. Thus, to linear order in
the strain, there is no restoring force to $xy$-stresses
\cite{stress_tensors}
\begin{align}
\sigma_{xy} = \frac{\partial f_{D_{2h}}^{\text{soft}}}{\partial
u_{xy}}  \, , \label{stress_1}
\end{align}
i.e., to opposing forces along $\pm \tilde{\ev}_x$ applied to
opposite surfaces with normal along $\pm \tilde{\ev}_y$ or
opposing forces along $\pm \tilde{\ev}_y$ applied to opposite
surfaces with normal along $\pm \tilde{\ev}_x$, see
Fig.~\ref{fig:biaxialSoftness}.

Note that there is an interesting parallel between the biaxial
smectics considered here and biaxial nematics formed spontaneously
from an isotropic elastomer. Warner and
Kutter~\cite{warner_kutter_2002} predicted that these biaxial
nematics have many soft modes, and one of these is identical to
the soft mode of biaxial smectics discussed above.  Finally, we
observe that a biaxial smectic has the same point-group symmetry
$D_{2h}$ as an orthorhombic crystal.  However, because it is
formed via spontaneous symmetry breaking from a state with
$D_{\infty h}$, unlike an equilibrium orthorhombic crystal, it
exhibits soft elasticity in the $xy$-plane.  An orthorhombic state
can be reached via a symmetry-breaking transition from a
tetragonal state, which exhibits square symmetry in the $xy$
plane. Rather than exhibiting the soft elasticity discussed above
such an orthorhombic system exhibits martinsitic elasticity
\cite{martensite} in which domains of different orientation are
produced in response to stress perpendicular to the stretch
direction in the $xy$ plane.

\subsection{Rotational invariance and soft extensional strains}
\label{biaxialRotationalInvariance} In addition to the softness
under $u_{xy}^\prime$ shears, the new phase is, provided that the
experimental boundary condition are right, soft with respect to
certain extensional strains. In the following paragraphs, we will
discuss this softness in some detail. The mechanism at work here
is intimately related to a mechanism leading to a similar softness
in nematic elastomers, and thus our reasoning will follow closely
well known arguments for nematic
elastomers~\cite{golubovic_lubensky_89,LubenskyXin2002}. The form
of these soft strains depends only on symmetry and the nature of
the broken-symmetry phase; it is not restricted to small strains
or to systems described by a Landau expansion of the free energy.
First, we will consider global rotations in the $xy$-plane of the
reference space of our biaxial elastomers. On the one hand this
will prepare the ground for understanding the softness of
extensional strains, and on the other hand, it will allow us to
understand the vanishing of the elastic constant $C_{xyxy}$ from a
somewhat different perspective. Then we calculate the energy cost
of soft extensional strains and finally we comment on experimental
implications.

We saw in Sec.~\ref{biaxialPhaseTransition} that the direction of
the spontaneous anisotropy $xy$-plane, or in other words, the
direction of the $c$-director $\tilde{\brm{c}}$, is arbitrary.
Thus, equilibrium states characterized, respectively, by
$\tens{u}^0$ and $\tens{O}_{R,z}  \tens{u}^0  \,
\tens{O}_{R,z}^{-1}$, where $\tens{O}_{R,z}$ describes an
arbitrary rotation in the reference space about the $z$ axis, must
have the same energy.  With
\begin{align}
\label{refRotz} \tens{O}_{R,z} = \left(
\begin{array}{ccc}
\cos \vartheta &  - \sin \vartheta & 0\\
 \sin \vartheta & \cos \vartheta & 0\\
0 & 0& 1
\end{array}
\right)
\end{align}
describing a {\em counterclockwise} rotation of vectors in the
reference space through $\vartheta$ about the $z$-axis, the strain
\begin{subequations}
\label{zeroStrain}
\begin{align}
\label{zeroStrainA} \tens{u}^\prime (\vartheta) &= \big(
\tens{\Lambda}^{0T} \big)^{-1} \big[ \tens{O}_{R,z} \, \tens{u}^0
\, \tens{O}_{R,z}^{-1} - \tens{u}^0 \big]  \big(
\tens{\Lambda}^{0} \big)^{-1}
\\
\label{zeroStrainB} &= \frac{r_\perp -1}{4} \left(
\begin{array}{ccc}
- r_\perp^{-1} ( 1- \cos 2\vartheta) & r_\perp^{-1/2} \sin 2\vartheta & 0\\
r_\perp^{-1/2} \sin 2\vartheta & 1- \cos 2\vartheta & 0\\
0 & 0& 0
\end{array}
\right)
\end{align}
\end{subequations}
must not cost any elastic energy. Our elastic energy
density~(\ref{biaxEn}) contains only second-order terms in
$u_{ij}^\prime$, and hence it can at best be invariant with
respect to infinitesimal rotations~\cite{footnoteSmallRotation}.
However, even for infinitesimal $\vartheta$, the strain
$\tens{u}^\prime (\vartheta)$ has nonzero components, namely
\begin{align}
u^\prime_{xy} (\vartheta) = u^\prime_{yx} (\vartheta) =
\frac{r_\perp -1 }{2 \sqrt{r_\perp}} \, \vartheta \, .
\end{align}
Thus, as it does in Eq.~(\ref{biaxEn}), the elastic constant
$C_{xyxy}$ of the term~(\ref{conventionalTerm}) must indeed
vanish, and this vanishing can be understood as a result of the
spontaneous symmetry breaking in the $xy$-plane.

The existence of zero-energy strains that reproduce rotations in
the reference space has consequences reaching further then just
the softness with respect to shear strains $u^\prime_{xy}$, viz.\
depending on the experimental boundary conditions, extensional
strains $u^\prime_{xx}$ and $u^\prime_{yy}$ can also be soft. If
the boundary conditions are such that no relaxation of strains is
allowed, then strains $u^\prime_{xx}$ and $u^\prime_{yy}$ will
cost an elastic energy proportional to $(u^\prime_{xx})^2$ and
$(u^\prime_{yy})^2$, respectively. If, however, strain relaxation
is allowed and one imposes for example $u^\prime_{yy}$ with the
right sign, then $u^\prime_{xx}$ and $u^\prime_{xy}$ can relax
under the right circumstances to produce the zero-energy strain of
Eq.~(\ref{zeroStrain}), i.e., to make $u^\prime_{yy}$ a soft
deformation.

To discuss this in more detail, let us assume for concreteness
that the anisotropy in the $xy$-plane is positive, $r_\perp > 1$.
Then it follows from Eq.~(\ref{zeroStrain}) that $u^\prime_{yy}$
is positive and that $u^\prime_{xx}$ is negative for soft strains
and that we consequently can only have soft elasticity for
$u^\prime_{yy} >0$ and $u^\prime_{xx}<0$. Let us consider here as
an example a strain with $u^\prime_{yy} >0$, i.e., a stretch of
the sample along $y$. Comparison with Eq.~(\ref{zeroStrain}) shows
that, if strain relaxation is allowed and $u^\prime_{xx}$ and
$u^\prime_{xy}$ relax to
\begin{subequations}
\begin{align}
u^\prime_{xx} &= - r_\perp^{-1}  u^\prime_{yy} \, ,
\\
u^\prime_{xy} &=  \pm\sqrt{\frac{u^\prime_{yy} (r_\perp - 1 -
2u^\prime_{yy})}{2r_\perp}} \, ,
\end{align}
\end{subequations}
 then $u^\prime_{yy}$ is converted into a zero-energy rotation through
an angle
\begin{align}
\label{effectiveAngle} \vartheta = \pm
\frac{1}{2}\sin^{-1}\left[\frac{2}{r_\perp - 1}\sqrt{2 u'_{yy}
(r_{\perp} - 1 - 2 u'_{yy})}\right]  .
\end{align}
Thus, $u^\prime_{yy}$ costs no elastic energy as long as $0
<u^\prime_{yy}<(r_\perp - 1)/2$.

When $u'_{yy}$ is increased from zero, $\vartheta$ increases from
zero until it reaches $\pi/2$ at $u'_{yy} = (r_\perp -1)/2$ [and
$u'_{xx} = (r_\perp^{-1} -1)/2$, $u^\prime_{xy} =0$] at which
point, the deformation tensor defined by $\Lambda'_{0ij}= \partial
R_i/\partial x'_j$ is
\begin{equation}
\tens{\Lambda}^\prime_{0}=  \left(
\begin{array}{ccc}
r_\perp^{-1/2} & 0 & 0 \\
0 & r_\perp^{1/2} & 0 \\
0 & 0 & 1
\end{array} \right) ,
\label{Lambda_0'}
\end{equation}
which leads via $\Lambda_{ij} = \Lambda_{ik}^0 \Lambda'_{0kj}$ to
an overall deformation
\begin{equation}
\tens{\Lambda} =  \left(
\begin{array}{ccc}
\Lambda_{yy}^0 & 0 & 0 \\
0 & \Lambda_{yy}^0 \sqrt{r_\perp} & 0 \\
0 & 0 & \Lambda_{zz}^0
\end{array}
\right)
\end{equation}
relative the original uniaxial state. Thus, at $\vartheta =
\pi/2$, $\Lambda_{ij}$ is identical to $\Lambda_{ij}^0$ except
with $\Lambda_{xx}^0= \sqrt{r_\perp}\Lambda_{yy}^0$ replaced by
$\Lambda^0_{yy}$ and $\Lambda_{yy}^0$ replaced by
$\Lambda_{xx}^0$, i.e., the $x$ and $y$ axes have been
interchanged in going from $\vartheta = 0$ to $\vartheta = \pi/2$.
In the process, $\tilde{\brm{c}}$ rotates from being parallel the
$x$-axis to being parallel to the $y$-axis \cite{c_rot}.

For $u_{yy}' >(r_\perp -1)/2$, there is no real solution for
$\vartheta$, and a further increase in $u_{yy}'$, measured by
$\delta u_{yy}^\prime = u_{yy}'- (r_\perp - 1 )/2$, which
stretches the system along the space-fixed $y$-axis, cost energy.
Since the anisotropy axis is now along the $y$- rather than the
$x$-axis, this stretching costs the same energy as it would have
cost to stretch the original system with anisotropy axis along the
space-fixed $x$-axis by the same amount. The $yy$-component of the
strain relative to the state with $\vartheta = \pi/2$ is $\Delta
u_{yy}' =( \Lambda_{0yy}')^{-2} \delta u_{yy}' = r_\perp^{-1}
\delta u_{yy}'$.  Thus, because the $x$- and $y$-axes have been
interchanged, the free energy as a function of $u_{yy}'$ is
\begin{align}
\label{biaxEnSoftStrecht} f_{D_{2h}}^{\text{soft}}  &= \left\{
\begin{array}{lcc}
0 & \mbox{for} & \delta u^\prime_{yy} < 0 \, ,
\\
\textstyle{\frac{1}{2 }} \, Y_x \, (\delta u^\prime_{yy})^2 &
\mbox{for} & \delta u^\prime_{yy} > 0 \, .
\end{array}
\right.
\end{align}
where
\begin{align}
Y_x & = \frac{1}{r_\perp^2} \, \bigg\{ C_{xxxx}
\nonumber \\
&- \frac{C_{yyyy} C_{zzxx}^2 + C_{zzzz} C_{xxyy}^2 - 2
C_{xxyy}C_{xxzz}C_{yyzz}}{C_{yyyy}C_{zzzz} - C_{yyzz}^2} \bigg\} .
\end{align}
is the is Young's modulus for stretching along the anisotropy axis
in the $xy$-plane (originally along $x$).

\begin{figure}
\centerline{\includegraphics[width=5.0cm]{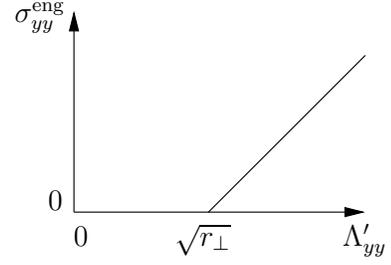}}
\caption{Schematic plot (arbitrary units) of the engineering
stress $\sigma^{\text{eng}}_{yy}$ versus the deformation
$\Lambda^\prime_{yy}$ for a soft biaxial elastomer with an
equilibrium order in the $xy$-plane along $x$ and positive
anisotropy, $r_\perp > 0$. Up to a critical deformation
$\Lambda^\prime_{0yy} = \sqrt{r_\perp}$ the sample responds to the
deformation merely by rotating $\tilde{\brm{c}}$ and consequently
the stress is zero. Above $\Lambda^\prime_{0yy}$ the sample
stretches along the new direction $y$ of the c-director and
$\sigma^{\text{eng}}_{yy}$ grows linearly for small $\Lambda_{yy}'
- \Lambda_{0yy}'$.} \label{fig:stressStrain}
\end{figure}

Equation~(\ref{biaxEnSoftStrecht}) has tangible implications for
the stress-strain behavior of soft biaxial elastomers. The stress
that is usually measured in experiments is the engineering stress,
i.~e., the force per unit area of the reference state. For the
extensional strain under discussion here, the engineering stress
\cite{stress_tensors} is to leading order
\begin{align}
\label{engineeringStress} \sigma^{\text{eng}}_{yy}  &=
\frac{\partial f_{D_{2h}}^{\text{soft}} }{\partial
\Lambda^\prime_{yy}} = \left\{
\begin{array}{lcc}
0 & \mbox{for} & \Lambda^\prime_{yy} < \sqrt{r_\perp}
\\
 Y_x \, \Lambda^\prime_{0yy} \, \delta u_{yy}'
& \mbox{for} & \Lambda^\prime_{yy} > \sqrt{r_\perp}
\end{array}
\right. .
\end{align}
For $\Lambda_{yy}'$ near $\Lambda_{0yy}'$, $\delta u_{yy}' \approx
\Lambda_{0yy}' (\Lambda_{yy}' - \Lambda_{0yy}')$, and
$\sigma^{\text{eng}}_{yy}\approx Y_x (\Lambda_{yy}' -
\Lambda_{0yy}')$. Figure~\ref{fig:stressStrain} depicts the
dependence of $\sigma^{\text{eng}}_{yy}$ on $\Lambda^\prime_{yy}$.
From $\Lambda^\prime_{yy} =0$ up to a critical deformation
$\Lambda^\prime_{0yy} = \sqrt{r_\perp}$ the stress is zero. Above
the critical deformation, $\sigma^{\text{eng}}_{yy}$ grows
linearly from zero.

\section{Smectic-$C$ elastomers -- strain-only theory}
\label{smecticElastomers1} In this section we use the strain-only
theory to study the phase transition from a uniaxial Sm$A$
elastomer to a Sm$C$ elastomer when $C_5$ becomes negative in
response to a Sm$C$-ordering of the mesogenic component.

\subsection{Phase transition from uniaxial to smectic-$C$ elastomers}
\label{smecticElastomers1Transition} When $C_5$ is driven
negative, the uniaxial state becomes unstable to shear in the
planes containing the anisotropy axis, and the uniaxial energy
(\ref{uniEn}) must be augmented with higher-order terms involving
$u_{az}$ to stabilize the system:
\begin{align}
\label{uniEn2} f_{\text{uni}}^{(2)} &= f_{\text{uni}} + D_1\,
u_{zz} u_{az}^2 + D_2 \,  u_{ii} u_{az}^2 + D_3 \, \hat{u}_{ab}
u_{az} u_{bz}
\nonumber \\
&+ E  \, (u_{az}^2)^2 ,
\end{align}
where we omit all unimportant symmetry-compatible higher-order
terms and we use $u_{zz}$ and $u_{ii}$ rather than $\eta_z$ and
$\eta$. To study the ordered phase of this free energy when
$C_5<0$, we proceed in much the same way as we did for the biaxial
state of $f_{\text{uni}}^{(1)}$. Using Eq.\ (\ref{uniEn}) for
$f_{\text{uni}}$, we complete squares to write
$f_{\text{uni}}^{(2)}$ as the sum
\begin{align}
f_{\text{uni}}^{(2)} = f_{\text{uni}}^{(2,1)} +
f_{\text{uni}}^{(2,2)} ,
\end{align}
of the two terms
\begin{subequations}
\begin{align}
\label{f21} f_{\text{uni}}^{(2,1)}& =  \textstyle{\frac{1}{2}} \,
C_1\, w_{zz}^2 + C_2 \, w_{ii}w_{zz} +\textstyle{\frac{1}{2}}\,
C_3 \, w_{ii}^2 + C_4 \, w_{ab}^2 \, ,
\\
\label{f22} f_{\text{uni}}^{(2,2)}& =  C_5 \, u_{az}^2 + E_R \,
(u_{az}^2)^2 \, ,
\end{align}
\end{subequations}
where $E_R$, which we assume to be positive, is a renormalized
version of $E$,
\begin{align}
\label{ER} E_R = E - \textstyle{\frac{1}{2}} \sigma^2\,  C_1 -
\sigma \tau \, C_2 - \textstyle{\frac{1}{2}} \tau^2 \, C_3 -
\textstyle{\frac{1}{2}} \omega^2 \, C_4 \, ,
\end{align}
and where
\begin{subequations}
\label{compositeC5}
\begin{align}
&w_{zz} = u_{zz} - \sigma u_{az}^2 \, ,
\\
&w_{ii} = u_{ii} - \tau u_{az}^2 \, ,
\\
&w_{ab} = \hat{u}_{ab} - \omega \left( u_{az}u_{bz} -
\textstyle{\frac{1}{2}}\delta_{ab} u_{cz}^2 \right) \, .
\end{align}
\end{subequations}
The coefficients in Eqs.~(\ref{ER}) and (\ref{compositeC5}) are
given by
\begin{align}
\label{resSigmaTau} \left(
\begin{array}{c}
\sigma
\\
\tau
\end{array}
\right) = \frac{-1}{C_1\, C_3 - C_2^2} \left(
\begin{array}{c}
C_3 D_1 - C_2 D_2
\\
C_1 D_2 - C_2 D_1
\end{array}
\right)
\end{align}
and
\begin{align}
\label{resOmega} \omega = - \frac{D_3}{2\, C_4} \, .
\end{align}
In the limit $C_3 \to \infty$ the coefficient $\tau$ vanishes
whereas $\sigma$, $\omega$, and $E_R$ remain finite.

The equilibrium value $u_{az}^0$ of $u_{az}$ is determined by
minimizing $f_{\text{uni}}^{(2,2)}$, which has $xy$ symmetry.
Provided that we choose our coordinate system so that its $x$-axis
is along the direction of ordering, the corresponding equation of
state,
\begin{align}
\label{EQSC5} C_5 \, u_{az}^0 + 2 \, E_R \, u_{az}^0 (u_{bz}^0)^2
= 0 ,
\end{align}
leads to
\begin{align}
u_{yz}^0 = 0 \quad \mbox{for} \quad C_5 >0 \quad \mbox{and} \quad
C_5 <0
\end{align}
and
\begin{align}
S \equiv u_{xz}^0 = \left\{
\begin{array}{ccc}
0& \quad \mbox{for} \quad & C_5 >0 \, ,
\\
\pm \sqrt{-C_5 / (2 E_R)}& \quad \mbox{for} \quad& C_5 <0 \, .
\end{array}
\right.
\end{align}
From Eqs.~(\ref{f21}) and (\ref{compositeC5}) the other components
of $\tens{u}^0$ follow as
\begin{subequations}
\begin{align}
& u_{xx}^0 = \textstyle{\frac{1}{2}} \, (\tau + \omega- \sigma)\,
S^2   \, ,
\\
&  u_{yy}^0 = \textstyle{\frac{1}{2}} \, (\tau- \omega -\sigma) \,
S^2 \, ,
\\
& u_{zz}^0 =  \sigma \, S^2  \, .
\end{align}
\end{subequations}
We learn that, unlike the biaxial case, $\tens{u}^0$ is not
diagonal; it has nonvanishing $xz$ and $zx$ components, which
implies $C_{2h}$ rather than $D_{2h}$ symmetry.

As already pointed out in Sec.~\ref{biaxialElastomers}, an
equilibrium strain tensor determines the corresponding equilibrium
deformation tensor only up to global rotations in the target
space. Here we choose our coordinate system in the target space so
that the transition from the $D_{2\infty}$ to the  $C_{2h}$ state
corresponds to a simple shear as shown in Fig.~\ref{fig1}, i.e.,
we choose the target space coordinates so that $\tan \phi =
\Lambda_{xz}^0/\Lambda_{zz}^0$ is nonzero but $\Lambda_{zx}^0 =
0$. Then the only nonzero components of $\tens{\Lambda}^0$ are
\begin{subequations}
\label{equiDefC5}
\begin{align}
\Lambda_{xx}^0 &= \sqrt{1 + 2 u_{xx}^0} = \sqrt{1 +  (\tau +
\omega- \sigma)\, S^2 } \, ,
\\
\Lambda_{yy}^0&=\sqrt{1+2 u_{yy}^0}  =   \sqrt{1 +  (\tau -
\omega- \sigma)\, S^2 } \, ,
\\
 \Lambda_{xz}^0 &= 2\,
\frac{u_{xz}^0}{\Lambda_{xx}^0} = \frac{2\, S}{\sqrt{1 +  (\tau +
\omega- \sigma)\, S^2 } }\, ,
\\
\Lambda_{zz}^0&=\sqrt{1 + 2 u_{zz}^0 - (\Lambda_{xz}^0)^2}
\nonumber \\
&= \sqrt{ 1 + 2\,  \frac{(\sigma -2)\, S^2 +  (\tau + \omega-
\sigma)\, \sigma\, S^4}{1 +  (\tau + \omega- \sigma)\, S^2}}\, .
\end{align}
\end{subequations}

\subsection{Elasticity of the smectic-$C$ phase}

Next we expand the elastic free energy density about the
equilibrium state. The expansion of $f_{\text{uni}}^{(2,1)}$
[Eq.~(\ref{f21})] is particularly simple and leads to
\begin{align}
\label{f21expanded} \delta f_{\text{uni}}^{(2,1)}& =
\textstyle{\frac{1}{2}} \, C_1\, (\delta w_{zz})^2 + C_2 \, \delta
w_{ii} \delta w_{zz} +\textstyle{\frac{1}{2}}\, C_3 \, (\delta
w_{ii})^2
\nonumber \\
&+ C_4 \, (\delta w_{ab})^2
\end{align}
with
\begin{subequations}
\begin{align}
& \delta w_{zz} = \delta u_{zz} - 2 \sigma S \, \delta u_{xz}  \,
,
\\
& \delta w_{ii}  =  \delta u_{ii} - 2 \tau S \, \delta u_{xz}  \,
,
\\
& \delta w_{xx} =   - \delta w_{yy} =  \textstyle{\frac{1}{2}} \,
( \delta u_{xx} - \delta u_{yy }) - \omega S \, \delta u_{xz}\, ,
\\
\label{yzEntering} & \delta w_{xy} = \delta w_{yx} = \delta u_{xy}
- \omega S\, \delta u_{yz} \, .
\end{align}
\end{subequations}
Expanding $f_{\text{uni}}^{(2,2)}$ [Eq.~(\ref{f22})], we find that
\begin{align}
\label{f22expanded} \delta f_{\text{uni}}^{(2,2)} = 4 E_R S^2
(\delta u_{xz})^2
\end{align}
is independent of $\delta u_{yz}$, and we might naively expect the
system to exhibit softness with respect to $u_{yz}$.  This,
however, is not the case because $f_{\text{uni}}^{(2,1)}$ depends
on $\delta u_{yz}$ via the relative strain~(\ref{yzEntering}).
Thus, the softness of the ordered phase with $C_{2h}$ symmetry is
more subtle than that of the biaxial phase with $D_{2h}$ symmetry,
as we will discuss in more detail further below. Assembling the
contributions~(\ref{f21expanded}) and (\ref{f22expanded}) we
obtain for the entire elastic free energy density to harmonic
order
\begin{widetext}
\begin{align}
\delta  f^{(2)} &=  2C_4 [\delta u_{xy}  - \omega S\,  \delta
u_{yz}]^2 + \textstyle{\frac{1}{2}} \, [C_1 + 2C_2 + C_3] \,
(\delta u_{zz})^2 +  4 E S^2 \, (\delta u_{xz})^2  +
\textstyle{\frac{1}{2}} \, [C_3 +C_4] \, \{ (\delta u_{xx})^2
+(\delta u_{yy})^2\}
\nonumber \\
&+ [C_2 + C_3] \delta u_{zz} \{ \delta u_{xx} + \delta u_{yy} \}+
[C_3 - C_4] \delta u_{xx} \delta u_{yy} + (2D_2 + D_3)S\,  \delta
u_{xx} \delta u_{xz} +  (2D_2 - D_3)S\,  \delta u_{yy} \delta
u_{xz}
\nonumber \\
&+  2(D_1 + D_2)S\,  \delta u_{zz} \delta u_{xz} \, .
\end{align}
\end{widetext}

Our remaining step in deriving the elastic free energy density of
Sm$C$ elastomers is to change from the variables of the old
uniaxial state to those of the new state by switching from $\delta
\tens{u}$ to the new strain tensor $\tens{u}^\prime$ as defined in
Eq.~(\ref{defuPrime}). With the equilibrium deformation tensor as
stated in Eqs.~(\ref{equiDefC5}) we obtain after some algebra
\begin{align}
\label{C2hEn} & f_{C_{2h}}^{\text{soft}} = \textstyle{\frac{1}{2}}
\, \bar{C} \left[ \cos \theta \, u^\prime_{xy} + \sin \theta \,
u^\prime_{yz} \right]^2 + \textstyle{\frac{1}{2}} \, C_{zzzz} \,
(u^\prime_{zz})^2
\nonumber \\
& + \textstyle{\frac{1}{2}} \, C_{xzxz} \, (u^\prime_{xz})^2 +
C_{zzxx} \, u^\prime_{zz} u^\prime_{xx} + C_{zzyy} \,
u^\prime_{zz} u^\prime_{yy}
\nonumber \\
& + \textstyle{\frac{1}{2}} \, C_{xxxx} \, (u^\prime_{xx})^2+
\textstyle{\frac{1}{2}} \, C_{yyyy} \, (u^\prime_{yy})^2 +
C_{xxyy} \, u^\prime_{xx} u^\prime_{yy}
\nonumber \\
& + C_{xxxz} \, u^\prime_{xx} u^\prime_{xz}+ C_{yyxz} \,
u^\prime_{yy} u^\prime_{xz} + C_{zzxz} \, u^\prime_{zz}
u^\prime_{xz} \, ,
\end{align}
where the angle $\theta$ and the elastic constants depend on the
original elastic constants in Eq.~(\ref{uniEn2}) and $S$ so that
one retrieves the uniaxial energy density~(\ref{uniEn}) for $S\to
0$. The explicit results for these quantities, which are somewhat
lengthy, can be found in Appendix~\ref{app:elasticConstants}. In
the incompressible limit, the specifics of $\theta$ and the
elastic constants get modified, however, without changing the form
of Eq.~(\ref{C2hEn}).

Having established the result~(\ref{C2hEn}), we are now in the
position to discuss the anticipated softness of the new state. Our
first observation is that the elastic energy density of
Eq.~(\ref{C2hEn}) has only 12 (including $\theta$) rather than the
13 independent elastic constants of conventional monolinic
solids~\cite{triclinic}. That is because here there are only two
rather than three independent elastic constants in the subspace
spanned by $u^\prime_{xy}$ and $u^\prime_{yz}$. Below, we present
two derivations of this result.

\begin{figure*}
\centerline{\includegraphics[width=12cm]{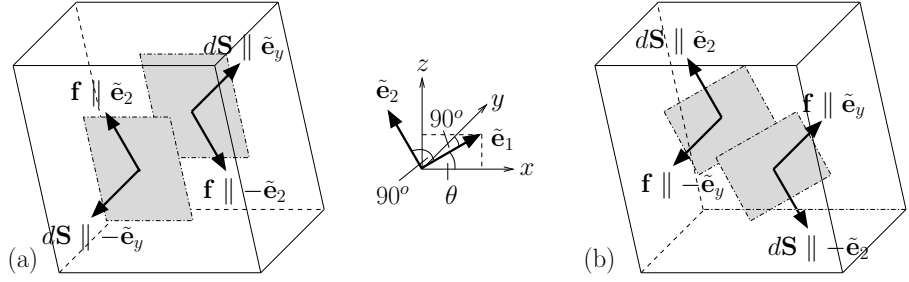}}
\caption{Soft distortions of Sm$C$ elastomers. There are no
restoring forces (a) for external forces in the $xz$-plane along
$\pm \tilde{\brm{e}}_2$ applied to opposing surfaces with normal
along $\pm \tilde{\brm{e}}_y$, and (b) for external forces  along
$\pm \tilde{\brm{e}}_y$ applied to surfaces with normal along $\pm
\tilde{\brm{e}}_2$.} \label{fig:stresses}
\end{figure*}

In the first derivation, we exploit the fact that $\cos \theta
u'_{xy} + \sin \theta u'_{yz}$ can be viewed as the dot product of
the ``vectors" $\vec{v}= (u'_{xy}, u'_{yz})$ and $\vec{e}_1 =
(\cos \theta, \sin \theta)$.  Thus, the first term in
Eq.~(\ref{C2hEn}),
\begin{align}
\textstyle{\frac{1}{2}} \, \bar{C} \left[ \cos \theta \,
u^\prime_{xy} + \sin \theta \, u^\prime_{yz} \right]^2 =
\textstyle{\frac{1}{2}} \, \bar{C} \, (\vec{e}_1 \cdot
\vec{v})^2\, ,
\end{align}
is independent of $\vec{e}_2 \cdot \vec{v}$, where
\begin{equation}
\vec{e}_2 = (-\sin\theta, \cos \theta)
\end{equation}
is the vector perpendicular to $\vec{e}_1$.  Thus, distortions of
the form $- \sin\theta u'_{xy} + \cos \theta u_{yz}=
\vec{e}_2\cdot \vec{v}$, i.e, distortions with $\vec{v}$ along
$\vec{e}_2$, cost no elastic energy. A manifestation of this
softness is that certain stresses cause no restring force and thus
lead to large deformations. To find these stresses we take the
derivative of the elastic energy density~(\ref{C2hEn}) with
respect to $u^\prime_{xy}$ and $u^\prime_{yz}$ which tells us that
\begin{align}
\label{softStress} -\sin \theta \, \sigma_{xy} + \cos \theta \,
\sigma_{zy} = \vec{e}_2 \cdot \vec{w} = 0 \, ,
\end{align}
where again, $\sigma_{ij}$ is the second Piola-Kirchhoff stress
tensor and where we have introduced the ``vector"
\begin{align}
\vec{w} = (\sigma_{xy}, \sigma_{zy})
\end{align}
in the $xz$-plane. Equation~(\ref{softStress}) means that there
are no restoring forces for external forces in the $xz$-plane
along $\pm \tilde{\brm{e}}_2 \equiv \pm (- \sin \theta, 0, \cos
\theta)$ applied to opposing surfaces with normal along $\pm
\tilde{\brm{e}}_y$, and there no restoring forces for external
forces along $\pm
 \tilde{\brm{e}}_y$ applied to surfaces with normal along $\pm  \tilde{\brm{e}}_2$.
These kinds of stresses are depicted in Fig.~\ref{fig:stresses}.

A second derivation of the softness of the Sm$C$ phase involves a
transformation to a rotated coordinate system, as described in
Eq.~(\ref{eq:basischange}), in which the first term in
Eq.~(\ref{C2hEn}) is diagonal. The components
$x^{\prime\prime}_{i'}$ of a reference-space vector $\xv^\prime$
expressed with respect to a rotated basis $\{
\tilde{\ev}_{i^\prime}^\prime | i^\prime = x^\prime, y^\prime,
z^\prime \}$, where
$\tilde{\ev}_{i^\prime}^\prime=O_{R,y;i'j}\tilde{\ev}_j$ with
\begin{align}
\label{refRoty} \tens{O}_{R,y} = \left(
\begin{array}{ccc}
\cos \theta &  0  & \sin \theta\\
0 & 1 & 0\\
- \sin \theta & 0& \cos \theta
\end{array}
\right)
\end{align}
describing a counterclockwise rotation of the reference-space basis
about the $y$-axis, are simply $x^{\prime\prime}_i =O_{R,y;ij} \,
x'_j$. The components of the strain matrix expressed in the
rotated basis are $u^{\prime \prime}_{i'j'} =
\tilde{\ev}_{i^\prime}^{\prime} \cdot \tens{u}' \cdot
\tilde{\ev}_{j^\prime}^{\prime}= O_{R,y;i'i}u_{ij}O_{R,y;jj'}^T$,
from which we obtain via Eq.~(\ref{baseRotU}) that $u^{\prime
\prime}_{x',y'} = \cos\theta \, u^\prime_{xy} + \sin \theta \,
u^\prime_{yz}$ and $u^{\prime \prime}_{y',z'} = -\sin\theta \,
u^\prime_{xy} + \cos \theta \, u^\prime_{yz}$. Thus, taking
$\theta$ to be the angle appearing in Eq.~(\ref{C2hEn}) and
dropping the double-prime from the strains, we obtain
\begin{align}
\label{C2hEnRotated} & f_{C_{2h}}^{\text{soft}} =
\textstyle{\frac{1}{2}} \, C_{x^\prime y^\prime x^\prime y^\prime}
\, (u_{x^\prime y^\prime})^2 + \textstyle{\frac{1}{2}} \,
C_{z^\prime z^\prime z^\prime z^\prime} \, (u_{z^\prime z^\prime
})^2
\nonumber \\
& + \textstyle{\frac{1}{2}} \, C_{x^\prime z^\prime x^\prime
z^\prime} \, (u_{x^\prime z^\prime})^2 + C_{z^\prime z^\prime
x^\prime x^\prime} \, u_{z^\prime z^\prime } u_{x^\prime x^\prime}
\nonumber \\
& + C_{z^\prime z^\prime y^\prime y^\prime } \, u_{z^\prime
z^\prime} u_{y^\prime y^\prime} + \textstyle{\frac{1}{2}} \,
C_{x^\prime x^\prime x^\prime x^\prime} \, (u_{x^\prime x^\prime
})^2
\nonumber \\
& + \textstyle{\frac{1}{2}} \, C_{y^\prime y^\prime y^\prime
y^\prime} \, (u_{y^\prime y^\prime })^2 + C_{x^\prime x^\prime
y^\prime y^\prime} \, u_{x^\prime x^\prime} u_{y^\prime y^\prime}
\nonumber \\
&+ C_{x^\prime x^\prime x^\prime z^\prime} \, u_{x^\prime
x^\prime} u_{x^\prime z^\prime}+ C_{y^\prime y^\prime x^\prime
z^\prime} \, u_{y^\prime y^\prime } u_{x^\prime z^\prime}
\nonumber \\
&+ C_{z^\prime z^\prime x^\prime z^\prime} \, u_{z^\prime
z^\prime} u_{x^\prime z^\prime} \, .
\end{align}
for the elastic energy density of Sm$C$ elastomers in the rotated
coordinates. Note that, in this coordinate system, the elastic
constants $C_{x^\prime y^\prime y^\prime z^\prime}$ and
$C_{y^\prime z^\prime y^\prime z^\prime}$ are zero, and the
elastic energy does not depend at all on $u_{y'z'}$. $C_{x^\prime
y^\prime x^\prime y^\prime}$ is identical to $\bar{C}$ and
$C_{y^\prime y^\prime y^\prime y^\prime} = C_{y y y y}$. The
remaining new elastic constants are non-vanishing conglomerates of
the elastic constants defined via Eq.~(\ref{C2hEn}) and sines and
cosines of $\theta$. We refrain here from stating further
specifics to save some space and because calculating these
specifics is a straightforward exercise.

The vanishing of $C_{x^\prime y^\prime y^\prime z^\prime}$ and
$C_{y^\prime z^\prime y^\prime z^\prime}$ means that Sm$C$
elastomers are soft with respect to shears in the $y^\prime
z^\prime$-plane. If one can cut a rectangular sample with faces
perpendicular to the $\tilde{\ev}_{i^\prime}^\prime$, then there
are no restoring forces for external forces along $\pm
\tilde{\ev}_{z^\prime}^\prime$ applied to opposing surfaces with
normal along $\pm \tilde{\ev}_{y^\prime}^\prime$, and for external
forces  along $\pm \tilde{\ev}_{y^\prime}^\prime$ applied to
surfaces with normal along $\pm \tilde{\ev}_{z^\prime}^\prime$.
This softness can be visualized as in
Fig.~\ref{fig:biaxialSoftness} with $\tilde{\ev}_x$ replaced by
$\tilde{\ev}_{z^\prime}^\prime$ and $\tilde{\ev}_y$ replaced by
$\tilde{\ev}_{y^\prime}^\prime = \tilde{\ev}_y$.

Before we go on, let us finally comment on the impact of the
softness on the phonon spectrum of Sm$C$ elastomers. A
comprehensive discussion of the phonon spectrum of course requires
a dynamical theory  of smectic elastomers. This is beyond the
scope of this paper and makes the topic of a separate
publication~\cite{stenull_lubensky_SmCdynamics}. Our current
theory allows us to address static phonons. To this end we switch
to Fourier space~\cite{fourierTransformation} and re-express the
elastic energy density~(\ref{C2hEn}) in terms of the Fourier transform
$\widetilde{\brm{u}}(\brm{q})$ with respect to $\xv^\prime$ of the
displacement field $\uv^\prime (\xv^\prime) = \Rv^\prime
(\xv^\prime) - \xv^\prime$. Neglecting the non-linear part of the
strains, cf.\ Eq.~(\ref{defStrain}), this yields a Fourier
transformed elastic energy density
$\widetilde{f}_{C_{2h}}^{\text{soft}}$ harmonic in
$\widetilde{\brm{u}}$ and the wavevector. With $\widetilde{u}_x =
\widetilde{u}_z = q_y =0$, $\widetilde{f}_{C_{2h}}^{\text{soft}}$
reduces to
\begin{align}
\widetilde{f}_{C_{2h}}^{\text{soft}} = \textstyle{\frac{1}{8}} \,
\bar{C} \, (\cos \theta \, q_x + \sin \theta \, q_z)^2 \,
\widetilde{u}_y^2 \, .
\end{align}
Hence, the energy cost is zero for phonon displacements
$\widetilde{\brm{u}} = (0, \widetilde{u}_y, 0)$ parallel to
$\tilde{\ev}_y$ with wavevector $\brm{q} \parallel
\tilde{\ev}_{z^\prime}$. By similar means one also finds that
there is no energy cost for $\widetilde{\brm{u}} \parallel
\tilde{\ev}_{z^\prime}$ with $\brm{q}
\parallel \tilde{\ev}_y$. This softness of these phonons has tangible
implications on the dynamics in that it leads, e.g., to a
vanishing of the corresponding sound
velocities~\cite{stenull_lubensky_SmCdynamics}.

An interesting question that we have left aside so far is, of
course, whether Sm$C$ elastomers are, like nematics or biaxial
smectics, soft with respect to certain extensional strains. We
will postpone this question to Sec.~\ref{smecticElastomers2} until
after we have developed our strain-and-director model for Sm$C$
elastomers. This will then allow us to discuss soft deformations
and strains more comprehensively including their impact on the
director.

\section{Smectic-$C$ elastomers -- theory with strain and director}
\label{smecticElastomers2} Our second theory for Sm$C$ elastomers,
to be presented in this section, explicitly accounts for the
smectic layers and for the director $\brm{n}$. It generalizes the
achiral limit of the continuum theory by Terentjev and
Warner~\cite{terentjev_warner_SmC_1994} in a formalism that
ensures invariance with respect to arbitrary rather than
infinitesimal rotations of both the director and mass points. We
will see as we move along that the properties of the transition to
the Sm$C$ phase predicted by this theory are identical to those of
the strain-only theory, discussed in the preceding section, in
which $C_5$ goes to zero.

\subsection{Reference- and target-space variables and the polar decomposition theorem}
\label{sec:polarDecomposition}

In traditional uncrosslinked liquid crystals, there is no
reference space, and all physical fields like the smectic
layer-displacement field $U$, the layer normal $\Nv$, and the
Frank director $\nv$ are defined at real, i.e., target-space
points $\Rv$, and they transform as scalars, vectors, and tensors
under rotations in the target space. In the Lagrangian theory of
elasticity, fields are defined at reference space points $\xv$,
and they transform into themselves under the symmetry operations
of that space. To develop a comprehensive theory of
liquid-crystalline elastomers, it is necessary to combine
target-space liquid crystalline fields and reference-space elastic
variables to produce scalars that are invariant under arbitrary
rotations in the target space and under symmetry operations of the
reference space. This requires that we be able to represent
vectors and tensors in either space~\cite{LubenskyXin2002}.

To be more specific, let $\bv$ be a target-space vector, which by
definition transforms under rotations to $\bv^{\prime} =
\tens{O}_T \bv$, and let $\tilde{\bv}$ be a reference-space
vector, which transforms to $\tilde{\bv}^{\prime} =
\tens{O}_R\tilde{\bv}$. Recall that both reference- and
target-space vectors exist in the same physical Euclidean space
$\cal E$.  Therefore, there must be a transformation that converts
a given reference-space vector to a target-space vector and vice
versa while preserving length.  This transformation is provided by
the deformation matrix $\tens{\Lambda}$ and the matrix polar
decomposition theorem~\cite{HornJoh1991}, which states that any non-singular
square matrix can be expressed as the product of a rotation matrix
and a symmetric matrix.  If $\tilde{\bv}$ is a reference-space
vector, then $\tens{\Lambda}\tilde{\bv}$ is a target-space vector
that transforms under $\tens{O}_T$ but does not change under
$\tens{O}_R$ because under $\xv\rightarrow \xv'=\tens{O}_R \xv$
and $\Rv\rightarrow \Rv'=\tens{O}_T \Rv$, $\tilde{\bv} \rightarrow
\tilde{\bv}'=\tens{O}_R \tilde{\bv}$, $\tens{\Lambda} \rightarrow
\tens{\Lambda}'=\tens{O}_T \tens{\Lambda}\tens{}O_R^{-1}$ and
$\tens{\Lambda}\tilde{\bv} \rightarrow \tens{\Lambda}'
\tilde{\bv}'=\tens{O}_T \tens{\Lambda}\tens{O}_R^{-1} \tens{O}_R
\tilde{\bv} = \tens{O}_T \tens{\Lambda} \tilde{\bv}$.  The
transformation $\tilde{\bv} \to \tens{\Lambda}\tilde{\bv}$,
however, does not preserve length. To construct a transformation
that does, we simply multiply $\tens{\Lambda}$ by the square root
of the metric tensor to produce
\begin{equation}
\tens{O} = \tens{\Lambda} \, \tens{g}^{-1/2} . \label{eq:O}
\end{equation}
This operator clearly satisfies $\tens{O}^T \tens{O}=\tens{O} \,
\tens{O}^T = \tens{\delta}$ and $\det \tens{O} = 1$, and it is
thus a length-preserving rotation matrix. Equation~(\ref{eq:O}),
which can be recast in the form $\tens{\Lambda} = \tens{O} \,
\tens{g}^{1/2}$ is simply a restatement of the polar decomposition
theorem because $\tens{g}$ is a symmetric matrix. To first order
in $\partial u_i/\partial x_j$, $O_{ij}$ reduces to the standard expression
for an infinitesimal local rotation of an elastic body through an angle
$\brm{\Omega} =\frac{1}{2} \brm{\nabla} \times \uv$,
\begin{align}
\label{ExpRotMatrix} O_{ij} = \delta_{ij} - \epsilon_{ijk} \,
\Omega_k + \cdots \, ,
\end{align}
where $\epsilon_{ijk}$ is the Levi-Civita tensor. Equipped with
$\tens{O}$ we can convert (or rotate) any reference-space vector
$\tilde{\bv}$ to a target space vector $\bv$ via
\begin{align}
\label{targetToRefernce} \bv=\tens{O} \cdot \tilde{\bv}
\end{align}
and a target-space vector to a reference space vector via
\begin{align}
\label{fromNoToTilde} \tilde{\bv}= \tens{O}^T\cdot \bv \, .
\end{align}
An alternative interpretation of the relation between $\bv$ and
$\tilde{\bv}$ follows from
\begin{equation}
\bv =b_i \ev_i = O_{ij}\tilde{b}_j \ev_i \equiv \tilde{b}_j \tv_j
,
\end{equation}
with
\begin{equation}
\tv_j = O_{ji}^T \ev_j = g_{jk}^{-1/2} \frac{\partial
\Rv}{\partial x_k} ,
\end{equation}
where we used $\Lambda_{kl}^T \ev_l = (\partial R_l/\partial x_k)
\ev_l = \partial \Rv/\partial x_k$.  The set of vectors $\{\tv_i|
i = x,y,z\}$ forms an orthonormal target-space basis in the tangent
space of the deformed medium (recall that $\partial \Rv/\partial
x_j$ is a tangent-space vector). Thus, $\tilde{b}_i$ represents
the components of the target-space vector $\bv$ relative to the
orthonormal tangent-space basis defined by $\tv_i$.

We can now apply this formalism to the Frank director in smectic
elastomers.  The familiar director $\nv$ is a target-space vector,
which we can represent as
\begin{align}
\label{nFrank} \nv\equiv (\brm{c}, n_z)\, , \quad n_z = \sqrt{1 -
c_a^2} \, ,
\end{align}
where $\brm{c}$ is the so-called c-director.  Contractions of the
components of $\nv$ with the stress strain tensor $u_{ij}$ do not
produce a scalar because $n_i$ and $u_{ij}$ transform under
different operators.  To create scalar contractions, we can
convert $\nv$ to a reference-space vector via $\tilde{\nv} =
\tens{O}^{-1} \nv$, where $\tens{O}$ is defined by
Eq.~(\ref{eq:O}), with
\begin{align}
\label{nTilde} \tilde{\nv}\equiv (\tilde{\brm{c}}, \tilde{n}_z)\,
, \quad \tilde{n}_z = \sqrt{1 - \tilde{c}_a^2} \, .
\end{align}
Combinations like $\tilde{n}_a u_{ab} \tilde{n}_b$ and
$\tilde{n}_a u_{az} \tilde{n}_z$ are now scalars.  When
linearized, these combinations reproduce those in the original de
Gennes theory \cite{deGennes1}. Linearized deviations of the
target-space director from its equilibrium $\brm{n}_0$ can be
expressed as $\delta \brm{n} = \brm{\omega} \times \brm{n}_0$,
where $\brm{\omega}$ is a rotation angle.  Then $\delta
\tilde{\brm{n}} = \tilde{\brm{c}} = (\brm{\omega}-
\brm{\Omega})\times \brm{n}_0$ and, for example, $\tilde{c}_a
u_{az}  \rightarrow u_{az}[(\brm{\omega}-\brm{\Omega})\times
\brm{n}_0]_a$.  Since we are interested in the transition from the
Sm$A$ to the Sm$C$ phase in which $\brm{n}$ undergoes a rotation
through a finite rather than an infinitesimal angle relative to
its equilibrium in the Sm$A$ phase, we need to use the full
nonlinear representation of rotation matrices.

As we discussed in Sec.~\ref{lagrangeElasticity}, the reference
space can be endowed with an orthonormal basis
$\{\tilde{\ev}_i\}$. This space is anisotropic, and we take
$\tilde{\ev}_z$ to be along the uniaxial anisotropy direction of
the Sm$A$ material \cite{e-n0}. Crosslinking in the Sm$A$ phase
freezes in an anisotropy direction in the elastic network and,
therefore, a general preference for the reference-space director
to align along $\tilde{\ev}_z$.  This preference, present in
elastomers crosslinked in the nematic as well as the Sm$A$ phase,
is distinct from the preference, which we will discuss shortly,
for the director to adopt a preferred angle relative to the layer
normal.

\begin{figure}
\centerline{\includegraphics[width=8cm]{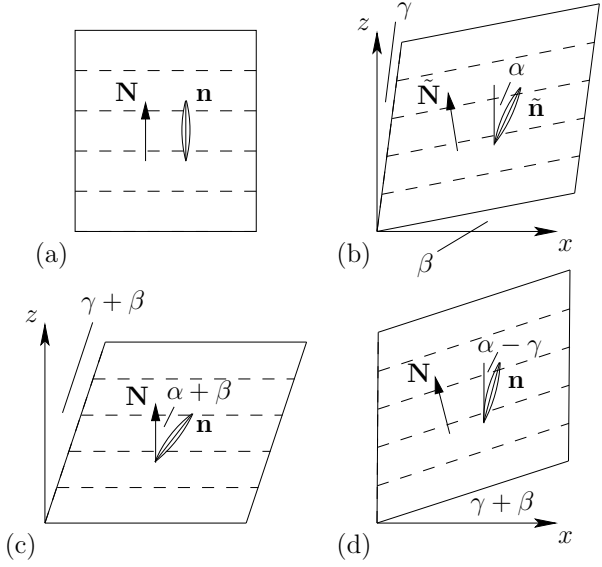}}
\caption{Schematic representation of distortions in the $xz$-plane
induced by the Sm$A$-to-Sm$C$ transition: (a) undistorted Sm$A$
phase, (b) Sm$C$ phase with a symmetric deformation tensor
$\tens{\Lambda}$, (c) Sm$C$ phase with $\Lambda_{xz}>0$ and
$\Lambda_{zx} = 0$, and (d) Sm$C$ phase with $\Lambda_{xz} = 0$
and $\Lambda_{zx} > 0$.} \label{fig:strains}
\end{figure}

An important property of $\tens{O}$ is that it reduces to the unit
matrix when $\tens{\Lambda}$ is symmetric, i.e., under pure shear
transformations, target- and reference-space vectors are
identical. Thus if a reference-space vector is known (calculated,
for example by minimizing a free energy that depends only on
reference-space vectors and tensors), the associated target-space
vector is obtained by rotating the reference-state vector by the
operator $\tens{O}$, which is the same operator that rotates the
pure shear configuration to the target-space configuration
described by $\tens{\Lambda}$. Figure~\ref{fig:strains} depicts
the effect on an initial reference state unit vector of a
symmetric shear and then a subsequent rotation to a final target
state.

\subsection{Development of a model for smectic elastomers}

Having established the relation between reference- and target
space vectors, we can now develop a complete phenomenological
energy density $f$ for smectic elastomers in terms of reference-space
variables only. For briefness, we will in the following often refer to energy densities simply as energies. We will be content with an expansion of $f$ in powers of the strain $u_{ij}$ and the $c$-director up to
fourth order. There are three distinct contributions to $f$: (1) the elastic energy $f_{\text net}$ of the anisotropic
network -- this is the harmonic energy of nematic elastomers
\cite{BlaTer94,WarnerTer2003} augmented by some nonlinear terms,
(2) the compression energy $f_{\text{layer}}$ of smectic layers,
and (3) the energy $f_{\text{tilt}}$ associated with tilt of the
nematic director relative to the layer normal.  The latter two
contributions are essentially the Chen-Lubensky (CL) model
\cite{Chen-Lubensky} model generalized to elastomers.  We will
assume that smectic layers are locked to the elastic network as is
the case when the network is crosslinked in the smectic phase.
The moduli and coupling constants in
$f_{\text{net}}$, $f_{\text{layer}}$, and $f_{\text{tilt}}$
contribute to the moduli and coupling constants in $f$.  We will
identify and calculate these contributions in what follows,
denoting each contribution by the appropriate superscript. Thus,
for example $C_1^{\text{net}}$ is the contribution of $C_1$ from
$f_{\text{net}}$.

The sum of all of the above contributions to the energy can
conveniently be decomposed as follows:
\begin{equation}
\label{completeEn} f= f_{\text{uni}} + f_{\text{nonlin}} + f_c +
f_{\text{coupl}} .
\end{equation}
We discuss each of these terms individually.  $f_{\text{uni}}$ is
the uniaxial elastic energy to quadratic order in strain of
Eq.~(\ref{uniEn}) with 5 elastic constants $C_1, ... ,C_5$.
Nonlinear strain energies are contained in $f_{\text{nonlin}}$.
This energy has many terms in general, but we will keep only those
terms that are relevant to the development of shear strain and
smectic-$C$ order:
\begin{equation}
f_{\text{nonlin}}= - B_1 \, u_{zz} u_{za}^2 + B_2 (u_{za}^2)^2 .
\end{equation}
To impose the full nonlinear incompressibility constraint to order $(u_{az}^2)$, we should include couplings of $u_{ii}$ to $u_{az}^2$.  If we did this, there would be a contribution to the energy of the form $\frac{1}{2}B[u_{ii} - 2u_{za}^2]^2$ where $B$ is the bulk compression modulus.  This would of course yield a contribution to $B_2$ of order $B$.  To treat such a term, we could replace $u_{ii}$ by $2 u_{za}^2 + \delta u_{ii}$ and express the energy in terms of $\delta u_{ii}$ rather than $u_{ii}$.  We will instead continue to consider the theory defined above without coupling between $u_{ii}$ and $u_{az}^2$.  The two theories effectively differ only in the value of $B_2$.
The energy associated with the development of $c$-order described
with nonzero ${\tilde c}_a$ is
\begin{equation}
\label{Htilt} f_c = \frac{r}{2} \, {\tilde c}_a^2 + \frac{v}{4} \,
({\tilde c}_a^2)^2 ,
\end{equation}
and the energy of director-strain coupling is
\begin{align}
\label{Hcoupl} f_{\text{coupl}} &=  \lambda_1\,  \tilde{c}_a^2
u_{zz} + \lambda_2 \, \tilde{c}_a^2 u_{ii}  +  \lambda_3 \,
\tilde{c}_a \hat{u}_{ab} \tilde{c}_b
\nonumber \\
&+  \lambda_4 \, \tilde{c}_a u_{az}  + \lambda_5 u_{zz} u_{az}
{\tilde c}_a .
\end{align}
Here we have retained terms linear in $u_{ij}$ and up to quadratic
order ${\tilde c}_a$.  We also keep the $\lambda_5$ term, which is
linear in ${\tilde c}_a$ and quadratic in $\tens{u}$; this
contributes to the behavior of Sm$A$ elastomers under uniaxial
extensional strain, which we will discuss in a separate paper
\cite{stenull_lubensky_smA_2006}.  The contributions of
$f_{\text{net}}$, $f_{\text{layer}}$, and $f_{\text{tilt}}$ to the
various terms in $f_{\text{uni}}$, $f_{\text{nonlin}}$, $f_c$, and
$f_{\text{coupl}}$ are summarized in Table~\ref{table1}.
\begin{table*}
\caption{\label{table1} Contributions to coefficients in $f$ from
$f_{\text{net}}$, $f_{\text{layer}}$ and $f_{\text{tilt}}$ with $a= \mu\,(p-1)^2/(2p)$.}
\begin{tabular}{|l|c|c|c|c|c|c|c|c|c|c|c|c|c|c|}
\hline
     &$C_1$ & $C_2$ & $C_3$ & $C_4$ & $C_5$ & $B_1$ & $B_2$
     &$\lambda_1 $& $\lambda_2$ & $\lambda_3$ & $\lambda_4$ &
     $\lambda_5$ & $r$ & $v$ \\
     \hline
  net & $3 \mu$ & $-\mu$ & $4B-\mu$ & $\mu$ & $\frac{1}{2} \mu
  \frac{(p+1)^2}{p}$ & $ a$ & $-a$ & $\mu \frac{2p^2-p-1}{2p}$ & $-\mu \frac{p -
  1}{2p}$ & $-\mu \frac{p - 1}{p}$ & $-\mu\frac{p^2-1}{p}$ & $-\frac{3}{2} a$ & $ \mu
  \frac{(p-1)^2}{p}$ & 0
  \\ \hline
  layer& $4B_{\text{sm}}$& 0 & 0 & 0 & 0 & $
  6 B_{\text{sm}}$ & $\frac{9}{2}B_{\text{sm}}$ & $2 B_{\text{sm}}$ & 0 & 0 & 0
  & $ 4 B_{\text{sm}}$ & 0 & $2 B_{\text{sm}}$
  \\ \hline
  tilt& 0 & 0 & 0 & 0 & $\frac{1}{2} r_t$ & $\frac{1}{2} r_t$ &
  $\frac{1}{2} r_t + \frac{1}{4} v_t$ & 0 & 0 & 0 & $r_t$ & $
  -\frac{1}{2} r_t$ & $ r_t $ & $v_t$
  \\ \hline
\end{tabular}
\end{table*}

We now discuss the individual contributions $f_{\text{net}}$,
$f_{\text{layer}}$, and $f_{\text{tilt}}$, beginning with
$f_{\text{net}}$. This is the energy of a nematic elastomer, which
we express in terms of reference-state variables only and which we
expand in powers of $u_{ij}$ and ${\tilde c}_a$. It can be
calculated from the neo-classical energy developed by Warner and
Terentjev \cite{WarnerTer2003} with an explicit volume compression
energy $\frac{1}{2} B (\det \tens{g}-1)^2$, where $B$ is the
compression modulus of order $3 \times 10^9$ Pa, in addition to
the entropic energy $\frac{1}{2} \mu \text{Tr} \tens{\Lambda}
\tens{\ell}_0 \tens{\Lambda}^T \tens{\ell}^{-1}$, where $\mu \sim
10^6$ Pa is the rubber shear modulus and $\tens{\ell}_0$ and
$\tens{\ell}$ are the polymer step-length tensors, respectively,
at sample preparation and in a general distorted state of the
system. The first step-length tensor is a reference-space tensor
with components
\begin{equation}
\ell_{0ij} = l_{\perp} [ \delta_{ij} + (p-1) {\tilde e}_i {\tilde
e}_j]\,  ,
\end{equation}
where ${\tilde \ev}\equiv {\tilde \ev}_z$ specifies the direction
of uniaxial anisotropy in the references state (denoted by $\nv_0$
in WT), $\ell_{\perp}$ is the step length perpendicular to
${\tilde \ev}$ and $p = \ell_{||}/\ell_{\perp}$, with $\ell_{||}$
the step length parallel to ${\tilde \ev}$. $\tens{\ell}^{-1}$ is
a target-space tensor with components $\ell^{-1}_{ij} =
\ell_{\perp}^{-1} [ \delta_{ij} + (p^{-1} -1) n_i n_j]$. Following
standard procedures \cite{WarnerTer2003}, $f_{\text{net}}$ can be
cast as the sum of a uniaxial energy the form of $f_{\text{uni}} +
f_{\text{nonlin}} + f_{\text{coupl}}$. Its contributions to the
various moduli and coupling constants are listed in table
\ref{table1}. The coefficients $C_5^{\text{net}}$,
$\lambda_4^{\text{net}}$ and $r^{\text{net}}$ satisfy
$C_5^{\text{net}} = (\lambda_1^{\text{net}})^2/(2 r^{\text{net}})$
as is required for soft nematic elastomers \cite{WarnerTer2003}.
We should logically add a semi-soft energy
\cite{FinKun97,VerWar96,War99} because we assume that the system
was crosslinked in the smectic phase. This will turn out to be
unnecessary because $f_{\text{layer}}$ and $f_{\text{tilt}}$
contribute the same kind of semi-soft terms but with greater
magnitude.

To derive both $f_{\text{layer}}$ and $f_{\text{tilt}}$, we need
to discuss in more detail the smectic displacement field $U$ and
the layer normal $\Nv$.  The smectic mass-density-wave amplitude
for a system with layer spacing $d$ has a phase
\begin{align}
\phi ( \Rv) = q_0 \, [R_z - U(\Rv)] \, ,
\end{align}
where $q_0 = 2 \pi /d$.  Since there is a one-to-one mapping from
the reference-space points $\xv$ to the target-space points
$\Rv(\xv)$, we can express $\phi$ as a function of $\xv$ as
\begin{align}
\phi(\xv) = q_0 \, [ z + u_z ( \xv) - U(\Rv(\xv))] \, .
\end{align}
We are only considering systems crosslinked in the smectic phase
in which the smectic mass-density wave cannot translate freely
relative to the reference material, and there is a term
\begin{align}
f_{\text{lock-in}} = \textstyle{\frac{1}{2}}\, A \, (u_z - U)^2
\end{align}
in the total free-energy density  that locks the smectic field $U$
to the displacement field $u_z$~\cite{lubensky&Co_94}. In what
follows, we will take this lock-in as given and set $U=u_z$. This
has some interesting consequences.  The smectic phase is now $\phi
= q_0 z$, which implies
\begin{equation}
\nabla_i \phi = \partial\phi/\partial R_i = q_0
[\Lambda^{-1}]_{zi} \, ,
\end{equation}
where we introduced the notation that $[M^{\alpha}]_{ij}$ is the
$ij$-component of the matrix $\tens{M}^{\alpha}$ for any matrix
$\tens{M}$ and exponent $\alpha$ (we retain the notation $M_{ij}
\equiv [M]_{ij})$. Thus, in the target space, the unit layer
normal reads
\begin{equation}
N_i = \frac{\nabla_i \phi}{|\bm{\nabla} \phi|} =
\frac{[\Lambda^{-1}]_{zi}}{[g^{-1}]_{zz}} \, .
\end{equation}
Using the polar decomposition theorem, we can calculate the
reference-space layer normal
\begin{equation}
{\tilde N}_i = \frac{[g^{-1/2}]_{iz}}{[g^{-1}]_{zz}^{1/2}} .
\end{equation}
With these definitions, we have
\begin{align}
{\tilde \nv} \cdot {\tilde \Nv} & = \nv \cdot \Nv \approx 1 -
\frac{1}{2} ({\tilde c}_a + u_{az} )^2 + \frac{1}{2} u_{zz}
{\tilde c}_a u_{az}
\nonumber \\
& + \frac{1}{2} u_{zz} u_{az}^2 - \frac{5}{8} (u_{az}^2 )^2  .
\end{align}
In this expression, we have retained the dominant terms necessary
to describe the Sm$A$-Sm$C$ transition and the Helfrich-Hurault
instabilities \cite{Helfrich-Hurault,buckling}  produced by an
extensional strain $u_{zz}$ along $z$.  We have not included
higher-order terms in ${\tilde c}_a$ and $u_{ij}$, which could
change the numerical values of our results but not their form.

The preferred spacing between smectic layers depends on the
orientation of the director relative to the layer normals.  If
$\nv$ is parallel to $\Nv$, the preferred spacing is $d$.  If
$\nv$ is not parallel to the smectic layer spacing should scale
approximately as $d \cos \Theta$, where $\Theta$ is the angle
between $\nv$ and $\Nv$.  A phenomenological energy that reflects
this preferences is
\begin{align}
f_{\text{layer}} & = \frac{1}{2} B_{\text{sm}} \, q_0^{-4} [(\nv
\cdot
{\mathbf \nabla} \phi )^2 - q_0^2 ]^2 \nonumber \\
& = \frac{1}{2} B_{\text{sm}} \left( [g^{-1}]_{zz} ({\tilde \nv} \cdot {\tilde
\Nv} )^2 - 1 \right )^2
\nonumber \\
& \approx 2 B_{\text{sm}} [u_{zz} + \textstyle{\frac{1}{2}}( {\tilde c}_a + u_{az} )^2 -
2 u_{az}^2 ] ^2 . \label{eq:flayer}
\end{align}
The smectic compression modulus $B_{\text{sm}}$ is of order $10^7$
Pa deep in the smectic phase though it vanishes as the transition
to the nematic phase is approached. $f_{\text{layer}}$ is the
generalization to elastomers of the compression energy in the
Chen-Lubensky \cite{Chen-Lubensky} model for Sm$A$ and Sm$C$
phases. We could have used the more isotropic compression energy
proportional to $[({\mathbf \nabla}\phi)^2 - q_0^2 ]^2$ instead.
It is the one studied by AW \cite{adams_warner_2005}. The
advantage of the CL energy over the latter energy is that it
encodes the tendency for layer spacing to decrease when $\Theta$
becomes nonzero.  If, for example, $u_{az} = 0$ and ${\tilde c}_a
\neq 0$, then to minimize $f_{\text{layer}}$, $u_{zz} = -
\frac{1}{2} {\tilde c}_a^2 <0$. Thus, as expected tilt decreases
$u_{zz}$ and layer spacing.  The CL energy, like the more
isotropic one, also has built in the physics of the
Helfrich-Hurault instability \cite{Helfrich-Hurault}, which we
will discuss in another paper \cite{stenull_lubensky_smA_2006}. If
${\tilde c}_a = 0$ and $u_{zz} \neq 0$, then $f_{\text{layer}}$ is
minimized when there is a shear strain, $u_{za} = \pm \sqrt{
u_{zz}/2}$.

Finally, we turn to the tilt energy.  This is most easily
expressed in terms of $\sin^2 \Theta = 1 - ({\tilde\nv} \cdot
{\tilde \Nv})^2$:
\begin{align}
f_{\text{tilt}}  &= \frac{1}{2} r_t \sin^2 \Theta + \frac{1}{4}
v_t
\sin^4 \Theta \nonumber \\
&\approx  \frac{1}{2} r_t[({\tilde c}_a + u_{az} )^2 - u_{zz}
u_{az} {\tilde c}_a - u_{zz} u_{az}^2 +  (u_{az}^2)^2 ]
\nonumber \\
&+  \frac{1}{4} v_t [({\tilde c}_a + u_{az})^2 ]^2 .
\end{align}
The modulus $r_t$ is generally of order but less than
$B_{\text{sm}}$. However it vanishes as the transition from the
Sm$A$ to the Sm$C$ phase in uncrosslinked smectics is approached.

\subsection{\label{sec:SmA-SmC}
Phase transition from smectic-$A$ to smectic-$C$ elastomers}

Having developed a full model for smectic elastomers that provides
a description of both the Sm$A$ and Sm$C$ phases, we can study the
transition from the Sm$A$ to the Sm$C$ phase.  In this transition
${\tilde c}_a$ becomes nonzero, and because of the coupling
between ${\tilde c}_a$ and $u_{az}$, $u_{az}$ also becomes
nonzero. Alternatively, we could say that the angle $\Theta
\approx {\tilde c}_a + u_{az}$ becomes nonzero and drives the
development of a nonzero $u_{az}$ because of a $\Theta-u_{az}$
coupling.  We will use the variables ${\tilde c}_a$ and $u_{az}$
to describe the Sm$A$-Sm$C$ transition. To keep our discussion
simple, we will focus on the development of $c$-order and include
only those terms in the free energy that play an important role in
this transition. Accordingly, we will ignore $f_{\text{nonlin}}$,
i.e., we set $B_1= B_2 = 0$), and we will set $\lambda_5 = 0$.
Setting these coefficients, which are relevant to the
Helfrich-Hurault instability, to zero does not lead to any
qualitative modification of our results for the Sm$A$-to-Sm$C$
transition. When $\lambda_1=\lambda_2= \lambda_3=0$, the model
described in Eqs.~(\ref{completeEn}) to (\ref{Hcoupl}) is
equivalent to that studied in
Ref.~\cite{terentjev_warner_SmC_1994} when polarization is
ignored. When $\tilde{c}_a$ is integrated out of $f$, the result
is identical to the elastic energy density $ f_{\text{uni}}^{(2)}$
of Sec.~\ref{smecticElastomers1} with $C_5$ renormalized to
$C_{5,R} = C_5 - \lambda_4^2/(2r)$, with $D_m$, $m=1,2,3$,
replaced by $\lambda_m \lambda_4^2/r$, and with $E$ replaced by
$(v/4 + r/2)\lambda_4^4/r^4$.

We can now analyze the transition to the Sm$C$ phase in exactly
the same way as we did in the strain only model of
Sec.~\ref{smecticElastomers1}. We complete the squares involving
the strains and the director-strain couplings. The resulting
elastic energy density is once more a sum of two terms,
\begin{align}
f = f^{(1)} + f^{(2)},
\end{align}
where
\begin{align}
\label{f1} f^{(1)}& =  \textstyle{\frac{1}{2}} \, C_1\,
\bar{w}_{zz}^2 + C_2 \, \bar{w}_{ii}\bar{w}_{zz}
+\textstyle{\frac{1}{2}}\, C_3 \, \bar{w}_{ii}^2 + C_4 \,
\bar{w}_{ab}^2 \, ,
\end{align}
is quadratic in the shifted strains
\begin{subequations}
\label{compositeCsmektic}
\begin{align}
&\bar{w}_{zz} = u_{zz} - \bar{\sigma} \, \tilde{c}_c^2 \, ,
\\
&\bar{w}_{ii} = u_{ii} - \bar{\tau} \, \tilde{c}_c^2 \, ,
\\
&\bar{w}_{ab} = \hat{u}_{ab} - \bar{\omega} \left( \tilde{c}_a
\tilde{c}_b - \textstyle{\frac{1}{2}}\delta_{ab} \tilde{c}_c^2
\right)  ,
\\
&\bar{w}_{az} = u_{az} - \bar{\rho} \, \tilde{c}_a \, ,
\end{align}
\end{subequations}
and where
\begin{align}
\label{f2} f^{(2)}& =   \textstyle{\frac{1}{2}} \, r_R \,
\tilde{c}_c^2 + \textstyle{\frac{1}{4}} \, v_R \, ( \tilde{c}_c^2
)^2 ,
\end{align}
depends on $\tilde{\brm{c}}$ only. The coefficients
$\bar{\sigma}$, $\bar{\tau}$, and $\bar{\omega}$ in
Eqs.~(\ref{compositeCsmektic}) are of the same form as the
coefficients $\sigma$, $\tau$, and $\omega$ of
Sec.~\ref{smecticElastomers1}, see Eqs.~(\ref{resSigmaTau}) and
(\ref{resOmega}), albeit with $D_l$, $l =1,2,3$, replaced by
$\lambda_l$. The coefficient $\bar{\rho}$ is given by
\begin{align}
\bar{\rho} = - \frac{\lambda_4}{2\, C_5} \, .
\end{align}
The renormalized elastic constants $r_R$ and $v_R$ in
Eq.~(\ref{f2}) read
\begin{subequations}
\label{renCoeffs}
\begin{align}
 r_R &= r - \frac{\lambda_4^2}{2\, C_5} \, ,
\\
v_R &= v - 2 \, \bar{\sigma}^2\,  C_1 - 4 \, \bar{\sigma}
\bar{\tau} \, C_2 - 2\,  \bar{\tau}^2 \, C_3
\nonumber \\
&- 2 \,  \bar{\omega}^2 \, C_4 + 4 \, \bar{\rho}^2 \, C_5 \, .
\end{align}
\end{subequations}
In the incompressible limit, the coefficient $\bar{\tau}$ vanishes
wheras the remaining coefficients and the renormalized elastic
constants $r_R$ and $v_R$ stay nonzero.

The transition to the Sm$C$ phase occurs at $r_R = 0$.  From Table
{\ref{table1}, we have $r= r_t + \mu(p-1)^2/p$, $\lambda_4 = r_t -
\mu (p^2 - 1)/p$, and $2 C_5 = r_t + \mu (p_1)^2/p$, and we find
that the critical value $r_t^c$ of $r_t$ at which the transition
occurs to be zero.  In other words, the coupling to the elastic
network does not affect the Sm$C$ transition temperature.  This
result is a direct consequence of the assumed semi-softness of the
elastomer in the absence of smectic ordering.

Next we minimize $f$ to assess the equilibrium states. With our
coordinate system chosen so that $\tilde{\brm{c}}$ aligns along
$x$, we obtain readily from Eq.~(\ref{f2}) that
\begin{align}
\tilde{c}_y^0 = 0 \quad \mbox{for} \quad r_R >0 \quad \mbox{and}
\quad r_R <0
\end{align}
and
\begin{align}
S \equiv \tilde{c}_x^0 = \sin \alpha = \left\{
\begin{array}{ccc}
0 & \quad \mbox{for} \quad & r_R >0 \, ,
\\
\pm \sqrt{-r_R / v_R} & \quad \mbox{for} \quad & r_R <0 \, ,
\end{array}
\right.
\end{align}
where $\alpha$ is the angle that the reference-space director
makes with the $z$-axis. The full reference space nematic director
is thus
\begin{equation}
\label{SmCrefSpaceDir} \tilde{\brm{n}} = (\sin \alpha, 0 , \cos
\alpha ) .
\end{equation}
Note that this corresponds to a counterclockwise rotation through
$\alpha$ about the $y$-axis of the original director
$\tilde{\brm{n}} = (0,0,1)$ in the Sm$A$ phase. The
director~(\ref{SmCrefSpaceDir})  is also the target space director
under a symmetric deformation tensor $\tens{\Lambda}^0$ as shown
in Fig.~\ref{fig:strains}(b).

The components of the equilibrium strain tensor then follow from
Eqs.~(\ref{f2}) as
\begin{subequations}
\begin{align}
& u_{xx}^0 = \textstyle{\frac{1}{2}} \, (\bar{\tau} +
\bar{\omega}- \bar{\sigma})\, S^2   \, ,
\\
&  u_{yy}^0 = \textstyle{\frac{1}{2}} \, (\bar{\tau}- \bar{\omega}
- \bar{\sigma}) \, S^2 \, ,
\\
& u_{zz}^0 =  \bar{\sigma} \, S^2  \, ,
\\
& u_{xz}^0 = u_{zx}^0 = \bar{\rho} \, S \, ,
\end{align}
\end{subequations}
and zero for the remaining components. Thus, to leading order in
the order parameter $S$, the equilibrium strain tensor has exactly
the same form as the one predicted by the strain-only theory of
Sec.~\ref{smecticElastomers1}. The only differences reside in the
specifics of the fore-factors of the $S$- and $S^2$-terms, which
are qualitatively unimportant.

Once again, we have to choose our coordinate system in target
space. As in Sec.~\ref{smecticElastomers1} we choose this system
so that the transition form Sm$A$ to Sm$C$ amounts to the simple
shear shown in Fig.~\ref{fig1}(c) with $\tan \, \phi =
\Lambda_{xz}^0/\Lambda_{zz}^0$, and $\Lambda_{zx}^0=0$. With this
choice,
\begin{equation}
\tens{\Lambda}^0 = \left(
\begin{array}{ccc}
\Lambda_{xx}^0 & 0 & \Lambda_{xz}^0 \\
0 & \Lambda_{yy}^0 & 0 \\
0 & 0 & \Lambda_{zz}^0
\end{array} \right),
\label{eq:L_simipleshear}
\end{equation}
 where
\begin{subequations}
\label{equiDefComplete}
\begin{align}
\Lambda_{xx}^0 &  =   \sqrt{1 +  (\bar{\tau} + \bar{\omega}-
\bar{\sigma})\, S^2 } \, ,
\\
\Lambda_{yy}^0 & =    \sqrt{1 +  (\bar{\tau} - \bar{\omega}-
\bar{\sigma})\, S^2 } \, ,
\\
 \Lambda_{xz}^0 & =   \frac{2\, \bar{\rho} \, S
\, }{\sqrt{1 +  (\bar{\tau} + \bar{\omega}- \bar{\sigma})\, S^2 }
}\, ,
\\
\Lambda_{zz}^0 & = \sqrt{1 + 2 \bar{\sigma}^2 S^2 + \frac{4
\bar{\rho}^2 S^2(1-S^2)}{1 + (\bar{\tau} + \bar{\omega}-
\bar{\sigma})\, S^2}}\, .
\end{align}
\end{subequations}

Knowing $\tilde{\brm{c}}^0$ and $\tens{\Lambda}^0$ we can discuss
what happens in the Sm$C$ phase to the layer normal, the director,
and the uniaxial anisotropy axis. Under the simple
shear~(\ref{equiDefComplete}), $(\Lambda_{zi}^0)^{-1} =
(\Lambda_{zz}^0)^{-1}\delta_{zi}$, and hence
\begin{equation}
\brm{N} = (0,0,1) .
\end{equation}
Thus, as expected, the shear deformation induced by the transition
to the Sm-$C$ phase slides the smectic layers parallel to each
other.  In this geometry, it does not rotate the layer normal.
Since $\brm{N}$ is parallel to the $z$-axis under simple shear,
the angle between the layer normal and the nematic director
$\brm{n}$ is the angle that the director makes with the $z$ axis
under simple shear.  This angle is simply $\Theta = \alpha+
\beta$, where $\beta$ is the angle through which the sample has to
be rotated to bring the symmetric-shear configuration to the
simple-shear configuration. Under symmetric shear, the symmetric
deformation tensor is given by
\begin{align}
\label{symmetricDeformation} \tens{\Lambda}_S =
\tens{g}^{1/2}=(\tens{\delta} + 2 \tens{u})^{1/2}
\end{align}
In order to calculate $\beta$, we need the symmetric equilibrium
deformation tensor $\tens{\Lambda}_S^0$, given by
Eq.~(\ref{symmetricDeformation}) with $\tens{g}$ replaced by
$\tens{g}^0 = \tens{\Lambda}^{0T} \tens{\Lambda}^0$. In terms of
the components of $\tens{\Lambda}_S^0$,
\begin{align}
\beta & = \tan^{-1} \left( [(1+ 2u^0)^{1/2}]_{zx}/[(1+2
u^0)^{1/2}]_{zz} \right)
\nonumber \\
& \approx u_{xz}^0  =2 {\overline \rho} S \approx \frac{p-1}{p+1}
S ,
\end{align}
where we replaced $r_t$ by $r_t^c = 0$ to obtain the final result.
Note that $u_{xz}$ and $\beta$ are positive as depicted in
Fig.~\ref{fig:strains}.  Tedious but straightforward algebra
verifies that the simple-shear deformation tensor
$\tens{\Lambda}^0$, whose components are given by
Eq.~(\ref{equiDefComplete}), satisfies $\tens{\Lambda}^0 =
\tens{O}_y (\beta ) (\tens{g}^0)^{1/2}$, where
\begin{align}
\label{rotationRefernceTarget} \tens{O}_y(\beta ) = \left(
\begin{array}{ccc}
\cos \beta & 0 & \sin \beta \\
0 & 1 & 0 \\
-\sin \beta & 0 & \cos \beta
\end{array}
\right)
\end{align}
is the matrix for a counter-clockwise rotation about the $y$-axis,
which is into the paper in Fig.~\ref{fig:strains}, through
$\beta$. The angle between $\nv$ and $\Nv$, which is equivalent to
the angle between $\nv$ and the $z$-axis, is
\begin{align}
\label{targetSpaceDirector} \Theta & = \alpha + \beta = \left(1 -
\frac{\lambda_4}{2 C_5} \right) S \approx \frac{2 p}{p+1} S .
\end{align}
The uniaxial anisotropy vector $\tilde{\ev}$ becomes $\brm{e} =
(\sin \beta, 0, \cos \beta )$.

Note finally that the angle $\gamma$ in Fig.~\ref{fig:strains} is
$\gamma= \tan^{-1}[(1+ 2 u^0)^{1/2}]_{xz}/[(1+ 2 u^0)^{1/2}]_{zz}
\approx u_{xz}$.  Thus, to lowest order $\gamma = \beta$. They
differ, however, at higher order in $S$. The tilt angle $\phi$
depicted in Fig.~\ref{fig1}(c) is given in terms of the angles
defined in Fig.~\ref{fig:strains} by $\phi = \beta + \gamma
\approx 2 \beta \approx 2 (p-1)/(p+1) S$. Thus, the spontaneous
mechanical tilt of the sample, as described by $\phi$, and the
tilt of the mesogens, as described by $\Theta$, are not equal.

\subsection{Elasticity of the smectic-$C$ phase}
\label{SmCelasticityII} To study the elastic properties of the
Sm$C$ phase we expand the elastic energy density $f$ about the
$C_{2h}$ equilibrium state. Expansion of $f^{(1)}$ to harmonic
order results in
\begin{align}
\label{f1expanded} \delta f^{(1)}& =  \textstyle{\frac{1}{2}} \,
C_1\, (\delta \bar{w}_{zz})^2 + C_2 \, \delta \bar{w}_{ii} \delta
\bar{w}_{zz} +\textstyle{\frac{1}{2}}\, C_3 \, (\delta
\bar{w}_{ii})^2
\nonumber \\
&+ C_4 \, (\delta \bar{w}_{ab})^2 + C_5 \, (\delta \bar{w}_{az})^2
\end{align}
with the composite strains
\begin{subequations}
\label{compStrainsCompleteModel}
\begin{align}
& \delta \bar{w}_{zz} = \delta u_{zz} - 2 \bar{\sigma} \, S \,
\delta \tilde{c}_x  \, ,
\\
& \delta \bar{w}_{ii}  =  \delta u_{ii} - 2 \bar{\tau} \, S \,
\delta \tilde{c}_x  \, ,
\\
& \delta \bar{w}_{xx} =   - \delta \bar{w}_{yy} =
\textstyle{\frac{1}{2}} \, ( \delta u_{xx} - \delta u_{yy }) -
\bar{\omega} S \, \delta \tilde{c}_x\, ,
\\
\label{yzEnteringCompleteModel} & \delta \bar{w}_{xy} = \delta
\bar{w}_{yx} = \delta u_{xy}  - \bar{\omega} \, S\,  \delta
\tilde{c}_y
\\
&\delta \bar{w}_{xz} = \delta \bar{w}_{zx} = \delta u_{xz} -
\bar{\rho} \left( 1- \textstyle{\frac{3}{2}}\, S^2 \right)
\delta\tilde{c}_x  \, ,
\\
&\delta \bar{w}_{yz} = \delta \bar{w}_{zy} = \delta u_{yz} -
\bar{\rho} \left( 1- \textstyle{\frac{1}{2}}\, S^2 \right) \delta
\tilde{c}_y  \, .
\end{align}
\end{subequations}
The expansion of $f^{(2)}$ is particularly simple. It leads to
\begin{align}
\label{f2expanded} \delta f^{(2)} =  v_R \, S^2 \, (\delta
\tilde{c}_x)^2 .
\end{align}
A glance at Eqs.~(\ref{f1expanded}),
(\ref{compStrainsCompleteModel}) and (\ref{f2expanded}) shows that
the 2 components of the c-director, $\delta\tilde{c}_x$ and
$\delta\tilde{c}_y$, play qualitatively different roles. Whereas
$\delta\tilde{c}_y$ appears only in the composite
strains~(\ref{compStrainsCompleteModel}), the component
$\delta\tilde{c}_x$ also appears in Eq.~(\ref{f2expanded}). In the
spirit of Landau theory of phase transitions, the term $v_R \, S^2
\, (\delta \tilde{c}_x)^2$ makes $\delta\tilde{c}_x$ a massive
variable. $\delta\tilde{c}_y$, on the other hand, is massless.
Since $\delta\tilde{c}_x$ is massive, the softness of the Sm$C$
phase that we expect from what we have learned in
Sec.~\ref{smecticElastomers1} cannot come from the relaxation of
$\delta\tilde{c}_x$. Rather it has to result from the relaxation
of $\delta\tilde{c}_y$. Anticipating this relaxation
$\delta\tilde{c}_y$, we rearrange $\delta f^{(2)}$ so that
$\delta\tilde{c}_y$ appears only in one place. Then we combine the
two contributions $\delta f^{(1)}$ and $\delta f^{(2)}$ and
integrate out the massive variable $\delta\tilde{c}_x$. Some
details on these steps are outlined in Appendix~\ref{app:steps}.

Our final step in deriving the elastic energy density is to change
from the strain variable $\delta \tens{u}$ to $\tens{u}'=
(\tens{\Lambda}^{0T})^{-1} \delta \tens{u} \,
(\tens{\Lambda}^{0})^{-1}$ with the equilibrium deformation tensor
as given in Eqs.~(\ref{equiDefComplete}). This takes us to
\begin{align}
\label{resBeforeRelaxation} f_{\text{Sm$C$}}& =
f_{C_{2h}}^{\text{soft}} +  \Delta \bigg[ \delta \tilde{c}_y
\nonumber\\
& +\Lambda_{yy}^0 \frac{ ( 2 \Lambda_{xx}^0  C_4  \Pi  +
\Lambda_{xz}^0  C_5  \Xi )  u_{xy}^\prime + \Lambda_{zz}^0 C_5
\Xi    u_{yz}^\prime }{\Delta} \bigg]^2,
\end{align}
where $\Pi = - \bar{\omega} \, S$, $\Xi = - \bar{\rho} (1-S^2/2)$,
and $\Delta = 2C_4 \Pi^2 + C_5 \Xi^2$. $f_{C_{2h}}^{\text{soft}}$
is exactly of the same form as the result stated in
Eq.~(\ref{C2hEn}). The only differences lie in the specifics of
the elastic constants. Our final formulas for the elastic
constants, which are rather lengthy, are collected in
App.~\ref{app:elasticConstants}.

Equation~(\ref{resBeforeRelaxation}) shows clearly that $\delta
\tilde{c}_y$ can relax locally to
\begin{align}
\label{resRelaxation} \delta \tilde{c}_y  = - \Lambda_{yy}^0
\frac{ ( 2 \Lambda_{xx}^0 \, C_4 \, \Pi  + \Lambda_{xz}^0 \, C_5
\, \Xi ) \,  u_{xy}^\prime + \Lambda_{zz}^0 C_5 \, \Xi  \,
u_{yz}^\prime }{\Delta}
\end{align}
which eliminates the dependence of the elastic energy density on
the linear combination of strains appearing on the right hand side
of Eq.~(\ref{resRelaxation}). In other words, the relaxation of
$\delta \tilde{c}_y$ produces an elastic energy density identical
to that of our strain-only model presented in
Sec.~\ref{smecticElastomers1},
\begin{align}
\label{resAfterRelaxation} f_{\text{Sm$C$}} =
f_{C_{2h}}^{\text{soft}} \,  ,
\end{align}
up to the aforementioned differences in the specific details of
the elastic constants. These details do not affect the elasticity
qualitatively. As in Sec.~\ref{smecticElastomers1}, the limit
$S\to 0$ reproduces the uniaxial elastic energy density of
Eq.~(\ref{uniEn}) and the incompressible limit leaves the form of
$f_{\text{Sm$C$}}$ unchanged. Most importantly, our model with
strain and director predicts the same softness of Sm$C$ elastomers
as our strain-only model of Sec.~\ref{smecticElastomers1}.

In our analysis we have completely neglected the Frank energy,
i.e., the effects of a non spatially homogeneous director.
However, it is legitimate to ask if it might affect the softness
of the material because, a priory, it is not impossible that the
Frank energy could lead to a mass for $\delta \tilde{c}_y$. We
address this question in appendix~\ref{app:frank}, where we find
that $\delta \tilde{c}_y$ remains massless even if the Frank
energy is included.

\subsection{Rotational invariance and soft extensional strains}
\label{smecticCRotationalInvariance} In this section we will
discuss the softness of Sm$C$ elastomers from the viewpoint of
rotational invariance in the $xy$-plane of the reference space.
The results presented here depend only on symmetries and not on
the detailed form of any free energy. They thus apply
quantitatively even when strains are large. Moreover, we will
inquire whether Sm$C$ elastomers are, like nematics and biaxial
smectics, soft with respect to certain extensional strains. Here,
we will use a somewhat different starting point than in
Sec.~\ref{biaxialRotationalInvariance} in that we first consider
soft deformations \cite{WarnerTer2003,Olmsted94,LubenskyXin2002}
rather than soft strains. Firstly, this is interesting in its own
right. Secondly, this will set the stage for a comparison of our
theory to the work of AW~\cite{adams_warner_SmC_2005} on the
softness of Sm$C$ elastomers. The results presented here depend
only on symmetry and the existence of a broken-symmetry state with
the symmetry of the Sm$C$ phase. They are not restricted to the
Landau expansion of the free-energy we used in preceding sections
or to small strains.

 Let us first determine the general
form of soft deformations. The equilibrium or ``ground state"
deformation tensor $\tens{\Lambda}^0$ maps points in the reference
space to points in the target space via $\brm{R} (\brm{x}) =
\tens{\Lambda}^0 \brm{x}$. Rotational invariance about the $z$
axis in the reference space ensures that $\brm{R}
(\tens{O}_{R,z}^{-1} \brm{x}) = \tens{\Lambda}^0
\tens{O}_{R,z}^{-1} \brm{x}$ describes a state with equal energy,
i.e., an alternative ground state. In other words, a deformation
described by
\begin{equation}
\overline{\tens{\Lambda}}^0 = \tens{\Lambda}^0 \tens{O}_{R,z}^{-1}
\end{equation}
has the same energy as one described by $\tens{\Lambda}^0$.  Any
deformation $\tens{\Lambda}$ relative to the original reference
system can be expressed in terms of a deformation
$\tens{\Lambda}'$ relative to the reference system obtained from
the original reference system via $\tens{\Lambda}^0$ through the
relation
\begin{equation}
\label{deformationArgument1} \tens{\Lambda} = \tens{\Lambda}'
\tens{\Lambda}^0 .
\end{equation}
Thus choosing $\tens{\Lambda}= \overline{\tens{\Lambda}}^0$, we
find that the deformation
\begin{equation}
\label{deformationArgument2} \tens{\Lambda}' = \tens{\Lambda}^0
\tens{O}_{R,z}^{-1} (\tens{\Lambda}^0)^{-1} \, ,
\end{equation}
with $\tens{O}_{R,z}$ the counterclockwise rotation matrix as
given in Eq.~(\ref{refRotz}), describes a zero-energy deformation
of the reference state represented by $\tens{\Lambda}^0$. Further
rotations of $\tens{\Lambda}'$ in the target space, of course, do
not change the energy, and the most general soft deformation
tensor is
\begin{equation}
\tilde{\tens{\Lambda}}' = \tens{O}_T \, \tens{\Lambda}'
\end{equation}
where $\tens{O}_T$ is an arbitrary target-space rotation matrix.

The reasoning just presented applies to any elastomer with
rotational invariance about the $z$-axis in reference space. Now
let us turn to the specifics for Sm$C$'s. Inserting the
equilibrium deformation tensor as calculated in
Sec.~\ref{smecticElastomers1Transition} or Sec.~\ref{sec:SmA-SmC},
we obtain
\begin{align}
\label{specificSoftDeformation} \tens{\Lambda}^\prime = \left(
\begin{array}{ccc}
\cos \vartheta &   r_\perp^{1/2} \sin \vartheta & s [1- \cos \vartheta] \\
-r_\perp^{- 1/2} \sin \vartheta & \cos \vartheta &  s \, r_\perp^{- 1/2} \sin \vartheta \\
0 & 0 & 1
\end{array}
 \right) ,
\end{align}
for Sm$C$ elastomers, where $s = \Lambda_{xz}^0/\Lambda_{zz}^0$
and, as before, $r_\perp = (\Lambda_{xx}^0/\Lambda_{yy}^0)^2$. Of
particular interest to our discussion of response to imposed
strain, which we present shortly, will be soft strains with a
vanishing $xy$ component. To construct such a soft deformation
tensor, we rotate through an angle $\omega$ about the $z$ axis,
\begin{align}
\label{softDeformationRoatedaboutZ} \tilde{\tens{\Lambda}}' =
\tens{O}_{T,z}(\omega) \, \tens{\Lambda}' .
\end{align}
Then, the condition $\tilde{\Lambda}'_{xy}=0$ is satisfied when
the target-space and reference space rotation angles are related
via $\tan \omega = r_{\perp}^{1/2} \tan \vartheta$ in which case
\begin{widetext}
\begin{equation}
\tilde{\tens{\Lambda}}^{\prime} = g(\vartheta) \left(
\begin{array}{ccc}
1 & 0 & - s [ 1- \cos \vartheta ] \\
\textstyle{\frac{1}{2}} r_{\perp}^{-1/2} (r_{\perp}-1 ) \sin 2
\vartheta& \quad \cos^2 \vartheta + r_{\perp} \sin^2 \vartheta &
\quad  \textstyle{\frac{1}{2}} r_{\perp}^{-1/2} s [-(r_{\perp} -1
) \sin 2
\vartheta + 2 r_\perp \sin \vartheta] \\
0 & 0 & 1
\end{array}
\right ) , \label{tilde_Lambda}
\end{equation}
\end{widetext}
where $g(\vartheta) = [1 + (r_{\perp} -1 ) \sin^2
\vartheta]^{-1/2}$. When $\vartheta = \pi/2$, then
\begin{equation}
\tilde{\tens{\Lambda}}'_0 \equiv \tilde{\tens{\Lambda}}'
(\vartheta = \pi/2)   = \left(
\begin{array}{ccc}
r_{\perp}^{-1/2} & 0 & - s r_{\perp}^{-1/2} \\
0 & r_{\perp}^{1/2} & s \\
0 & 0 & 1
\end{array}
\right), \label{Lambda_pi/2}
\end{equation}
which corresponds to an overall deformation
\begin{equation}
\tens{\Lambda} = \tilde{\tens{\Lambda}}'_0 \, \tens{\Lambda}^0 =
\left(
\begin{array}{ccc}
\Lambda_{yy}^0 & 0 & 0 \\
0 & \Lambda_{xx}^0 & \Lambda_{xz}^0 \\
0 & 0 & \Lambda_{zz}^0
\end{array}
\right) ,
\end{equation}
 i.e., to a shear deformation of
the original Sm$A$ in the $yz$- rather than the $xz$-plane.
Figure~\ref{fig:softRotation} shows the effect of deformations
$\tilde{\tens{\Lambda}}'$ for a series of values of $\vartheta$
between $0$ and $\pi/2$.
\begin{figure*}
\centerline{\includegraphics[width=10.5cm]{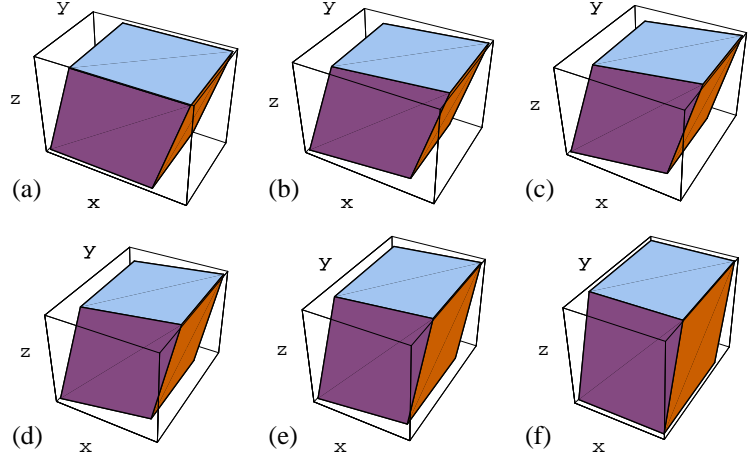}}
\caption{Effect of deformations $\tilde{\tens{\Lambda}}'$ as given
in Eq.~(\ref{tilde_Lambda}) for a series of values of $\vartheta$
between (a) $0$ and (f) $\pi/2$ with $\vartheta$ increased in
steps of $\pi/10$. In the process a parallelepiped-shaped sample
with initial shear in the $xz$-plane is transformed into a
parallelepiped with shear in the $yz$-plane that appears as if the
original parallelepiped had been rotated by $\pi/2$ about the
$z$-axis.} \label{fig:softRotation}
\end{figure*}

Knowing the deformation tensor $\tilde{\tens{\Lambda}}^{\prime}$,
we know how the shape of a sample changes in a soft deformation.
An equally interesting question is, however, how the orientation
of the mesogens changes under such a deformation. To address this
question, we recall from Sec.~\ref{sec:polarDecomposition} that
rotations in the reference space do not affect the target space
vectors such as the director $\nv$. Thus, $\nv$ does not change in
response to rotation described by $O_{R,z} ( \vartheta)$.  Of
course, under target space rotations, $\nv$ does change according
to $n'_i = O_{T, ij} n_j$. Therefore, the director, as given in
Eq.~(\ref{targetSpaceDirector}), changes in the process of the
soft deformation $\tilde{\tens{\Lambda}}^{\prime}$ to
\begin{align}
\label{changeInDirector} \brm{n} =  (\cos \omega \sin \alpha_T,
\sin \omega \sin \alpha_T, \cos \alpha_T) .
\end{align}

At this point we pause briefly to compare our findings to AW. Our
soft deformation tensor~(\ref{softDeformationRoatedaboutZ}) is, up
to differences in notation, identical to the soft deformation
tensor found by AW, see the last equation in the appendix
Ref.~\cite{adams_warner_SmC_2005}. The same holds true for the
change in the director associated with these soft deformations,
Eq.~(\ref{changeInDirector}). However, whereas the AW derivation
emphasizes geometric constraints, ours emphasizes that softness
arises from invariances with respect to reference space rotations
and the independent nature of reference and target space
rotations.  It should be noted that unlike the AW derivation, ours
does not impose incompressibility; rather the incompressibility
condition of the soft deformation arises naturally from the from
of Eq.~(\ref{deformationArgument2}).

To discuss the implication of the rotational invariance on the
Lagrange elastic energy, we now we switch from deformations to
strains. In terms of the soft deformation, the general form of the
soft strain tensor is given by
\begin{align}
\label{softStrainInTermsOfSoftDeformation} \tens{u}^\prime =
\textstyle{\frac{1}{2}} \big[ (\tens{\Lambda}^\prime)^T
\tens{\Lambda}^\prime - \tens{\delta} \big] =
\textstyle{\frac{1}{2}} \big[ (\tilde{\tens{\Lambda}}^\prime)^T
\tilde{\tens{\Lambda}}^\prime - \tens{\delta} \big]\,
\end{align}
independent of target-space rotations. Inserting
Eq.~(\ref{specificSoftDeformation}) or (\ref{tilde_Lambda}), it is
straightforward to check that
Eq.~(\ref{softStrainInTermsOfSoftDeformation}) and
Eq.~(\ref{zeroStrainA}) are equivalent. Using
Eq.~(\ref{specificSoftDeformation}) we find
\begin{subequations}
\begin{align}
\label{specificSoftStrain} u^\prime_{xx} (\vartheta)&= -
\frac{r_\perp -1}{4 \, r_\perp } \, [ 1 - \cos 2 \vartheta ] \, ,
\\
u^\prime_{xy}  (\vartheta) &=  \frac{r_\perp -1}{4 \,
\sqrt{r_\perp} } \, \sin 2 \vartheta \, ,
\\
u^\prime_{xz}  (\vartheta) & = -  \frac{s}{r_\perp} \, \left[ 1 -
(r_\perp -1) \cos \vartheta \right] \sin^2 \frac{\vartheta}{2} \,
,
\\
u^\prime_{yy}  (\vartheta) &=  \frac{r_\perp -1}{4 \, r_\perp } \,
[ 1 - \cos 2 \vartheta ] \, ,
\\
u^\prime_{yz}  (\vartheta) & = \frac{s}{ 2 \, \sqrt{r_\perp}}\,
\left[ r_\perp -  (r_\perp -1) \cos \vartheta \right] \sin
\vartheta  \, ,
\\
u^\prime_{zz}  (\vartheta) & = -  \frac{s^2}{r_\perp} \, \left[ 1
+ r_\perp  -  (r_\perp -1) \cos \vartheta \right] \sin^2
\frac{\vartheta}{2}
\end{align}
\end{subequations}
for the specifics of the soft strain.
Equation~(\ref{specificSoftStrain}) implies that the soft strain
has nonzero components for infinitesimal $\vartheta$, viz.\
\begin{subequations}
\begin{align}
\label{nonzeroSoftComponents} u^\prime_{xy}  (\vartheta) &=
u^\prime_{yx}  (\vartheta)  = \frac{r_\perp -1}{2 \,
\sqrt{r_\perp} } \,  \vartheta \, ,
\\
u^\prime_{yz}  (\vartheta) & = u^\prime_{zy}  (\vartheta) =
\frac{s}{ 2 \, \sqrt{r_\perp}}\,   \vartheta  \, .
\end{align}
\end{subequations}
To ensure that these infinitesimal strains do not cost elastic
energy the following combination of elastic constants has to
vanish:
\begin{align}
\label{vanishingCombination} C_{xyxy} \, (r_\perp -1)^2 + 2 \,
C_{xyyz} \, s (r_\perp -1)   + C_{yzyz} \, s^2 = 0 \, .
\end{align}
This equation is fulfilled if
\begin{subequations}
\label{SmCrelationsBetweenCs}
\begin{align}
 C_{xyxy}  &= \bar{C} \, \cos^2 \theta \, ,
 \\
 C_{xyyz}  &= \bar{C} \, \cos \theta \sin \theta \, ,
 \\
 C_{yzyz}  &= \bar{C} \, \sin^2 \theta \, ,
\end{align}
\end{subequations}
with the angle $\theta$ given by
\begin{align}
\label{valueForTheta} \theta = \tan^{-1} \left( \frac{1 -
r_\perp}{s} \right) .
\end{align}
Note by comparing Eqs.~(\ref{SmCrelationsBetweenCs}) and
(\ref{C2hEn}) that the analysis of the rotational invariance
presented here gives exactly the same relations between the
elastic constants as our analyses of the Sm$A$-to-Sm$C$ phase
transition presented in Sec.~\ref{smecticElastomers1} and
Sec.~\ref{sec:SmA-SmC} to \ref{SmCelasticityII}.

Modifying our arguments slightly, we can also understand from them
the vanishing of the elastic constants $C_{x^\prime y^\prime
y^\prime z^\prime}$ and $C_{y^\prime z^\prime y^\prime z^\prime}$
in the elastic energy density (\ref{C2hEnRotated}). Rotating the
soft strain tensor with the rotation matrix~(\ref{refRoty}) with
the rotation angle $\theta$ given by Eq.~(\ref{valueForTheta})
leads to a soft strain tensor that has in the limit of small
$\vartheta$ only
\begin{align}
u_{y^\prime z^\prime}  (\vartheta) & = u_{y^\prime z^\prime}
(\vartheta) = \frac{s}{ 2 \, \sqrt{r_\perp}}\,   \sqrt{1 +
\frac{(r_\perp -1)^2}{s^2}} \,  \vartheta
\end{align}
as nonzero components. For this strain to cost no energy
$C_{x^\prime y^\prime y^\prime z^\prime}$ and $C_{y^\prime
z^\prime y^\prime z^\prime}$ must be zero as they are in
Eq.~(\ref{C2hEnRotated}).

Next we turn to the question whether extensional strains can be
soft in Sm$C$ elastomers. As we did for biaxial smectics we
consider extensional strains along the $y$-axis, $u_{yy}^\prime >
0$, as a specific example. Again we assume, for the sake of the
argument, positive anisotropy in the $xy$-plane, $r_\perp > 1$.
Equations~(\ref{specificSoftStrain}) imply that $u_{yy}^\prime$ is
converted into a zero-energy rotation through an angle $\vartheta$
as given in Eq.~(\ref{effectiveAngle}), if the remaining
components relax to
\begin{subequations}
\begin{align}
u^\prime_{xx} &= - r_\perp^{-1}  u^\prime_{yy} \, ,
\\
u^\prime_{xy} &=  \sqrt{\frac{u^\prime_{yy} (r_\perp - 1 -
2u^\prime_{yy})}{2r_\perp}} \, ,
\\
u^\prime_{xz} &= - \frac{s}{2} \, \left[  1 - \frac{2\,
u^\prime_{yy} }{r_\perp} - \sqrt{1 - \frac{2\,
u^\prime_{yy}}{r_\perp -1}} \right] ,
\\
u^\prime_{yz} &= \frac{s}{2} \, \sqrt{\frac{2\, u^\prime_{yy}
}{r_\perp \, (r_\perp - 1)}}
\nonumber\\
&\times \left[ r_\perp - \sqrt{(r_\perp - 1)(r_\perp - 1 - 2 \,
u^\prime_{yy})} \right]
\\
u^\prime_{zz} &= s^2 \, \left[  1 - \frac{u^\prime_{yy} }{r_\perp}
- \sqrt{1 - \frac{2\, u^\prime_{yy}}{r_\perp -1}} \right] .
\end{align}
\end{subequations}
When $u_{yy}^\prime$ is increased from zero to $u_{yy}^\prime =
u_{yy}^c \equiv (r_\perp - 1)/2$, $\vartheta$ grows from zero to
$\pi/2$ and the state of the elastomer, originally described by
the equilibrium strain tensor $\tens{u}^0$ is changed without
costing elastic energy to
\begin{align}
\label{newEquiStrainTens} \tens{u}^\prime_0 = \frac{1}{2}  \left(
\begin{array}{ccc}
r_\perp^{-1} - 1 & 0 & - s / r_\perp \\
0  & r_\perp - 1  &  s \sqrt{r_\perp}\\
- s / r_\perp & s \sqrt{r_\perp} & s^2 (r_\perp^{-1} + 1)
\end{array}
 \right) ,
\end{align}
which is, of course, the strain tensor associated with the
deformation tensor of Eq.~(\ref{Lambda_pi/2}).

In this process, the shape of the sample changes as depicted in
Fig.~\ref{fig:softRotation}. As already discussed, the
configuration at $\vartheta = \pi/2$ describes a sample in which
the Sm$A$ phase was sheared in the $yz$- rather than the $xz$
plane.  Thus further increase in $u'_{yy}$ beyond $u_{yy}^c$ is
equivalent to increasing $u'_{xx}$ beyond zero in the original
sample sheared in the $xz$-plane.  Thus, the second
Piola-Kirchhoff stress is
\begin{equation}
\label{SmCsecondPiolaKirchStress} \sigma^{\text{II}}_{yy}  =
  \begin{cases}
  0 & \text{if $u_{yy}' < u_{yy}^c$}, \\
  Y_{x} \, (u_{yy}' - u_{yy}^c ) & \text{if $u_{yy}' >
u_{yy}^c$} ,\cr
  \end{cases}
\end{equation}
where $Y_{x}$ is the Youngs modulus for stretching along $x$.
Equation~(\ref{SmCsecondPiolaKirchStress}) implies that the stress
usually measured in experiments, i.e.\ the engineering stress, is
given at leading order by
\begin{equation}
\sigma^{\text{eng}}_{yy} =
  \begin{cases}
  0 & \text{if $\tilde{\Lambda}^\prime_{yy} < \sqrt{r_\perp}$}, \\
  Y_{x} \, \tilde{\Lambda}^\prime_{0yy}\,  ( \tilde{\Lambda}^\prime_{yy}-  \tilde{\Lambda}^\prime_{0yy} ) & \text{if $\tilde{\Lambda}^\prime_{yy} >
\sqrt{r_\perp}$} .\cr
  \end{cases}
\end{equation}
Therefore, when plotted as a function of the deformation
$\tilde{\Lambda}^\prime_{yy}$, the engineering stress
$\sigma^{\text{eng}}_{yy}$ for a Sm$C$ elastomer looks
qualitatively the same as the corresponding curve for a  nematic
or a biaxial smectic elastomer, cf.\ Fig.~\ref{fig:stressStrain}.

\section{Concluding remarks}
\label{concludingRemarks} In summary, we have presented models for
transitions from uniaxial Sm$A$ elastomers to biaxial and Sm$C$
elastomers: Landau-like phenomenological models
as functions of the Cauchy--Saint-Laurent strain tensor for both the transitions as well as a detailed model for the transition from the Sm$A$ to the smectic-$C$ phase. The detailed model includes contributions from the elastic network, smectic layer compression, and coupling of the Frank director to the smectic layer normal, and allowed for estimating the magnitudes of its phenomenological coupling constants. We employed the three models to investigate the nature of the soft
elasticity, required by symmetry, of monodomain samples of the
biaxial and Sm$C$ phases. 

We learned that biaxial smectic elastomers are soft with respect
to shears in the smectic plane. In addition to that we saw that
extensional strains can be converted by the material into
zero-energy rotations, provided the experimental boundary
conditions are not too restrictive and allow the remaining strain
degrees of freedom to relax. We illustrated this softness by
explicitly considering an elongation in the $y$ direction (the
direction perpendicular to the order in the smectic plane) as a
specific example. However, we could impose an extensional strain
in any direction and we would find softness, provided the
c-director has freedom to rotate into that direction. This
excludes only stretches in directions lying in the plane spanned
by the equilibrium c-director  and the initial uniaxial direction
(i.e., in our convention stretches in directions lying in the
$xz$-plane). Of course, the width of the soft plateau in the
stress-strain curve, c.f.~Fig.~\ref{fig:stressStrain}, depends on
how much the c-director can rotate until it has reached the
direction of a stretch. Therefore, the soft plateau will be most
pronounced for stretches perpendicular to the plane spanned by the
equilibrium c-director and the initial uniaxial direction (our
$y$-direction).

The softness of Sm$C$ elastomers is more intricate than that of
biaxial smectics. At first sight it seems as if it takes a very
specific combination of shears to achieve a soft response.
However, with the coordinate system chosen appropriately, it turns
out that Sm$C$ elastomers are soft with respect to certain
conventional shear strains (with our conventions shears in the
$y^\prime z^\prime$-plane). Even more important, as far as
possible experimental realizations of softness Sm$C$ elastomers
are concerned, is that these materials are also soft under
extensional strains. What we have said above for biaxial smectics
also applies here: the experimental boundary conditions must be
right and the direction of the imposed stretch must be so that the
c-director can rotate. 

As pointed out in the introduction, very recently considerable
experimental progress was made by Hiraoka {\em et
al}.~\cite{hiraoka&CO_2005}, who synthesized a monodomain sample
of a Sm$C$ elastomer forming spontaneously from a Sm$A$ phase upon
cooling. This is exactly the type of elastomer for which our Sm$C$
theory was made. Therefore, it seems well founded to hope that our
predictions for Sm$C$ elastomers can be tested experimentally in
the near future.

\begin{acknowledgments}
Support by the National Science Foundation under grant DMR 0404670
(TCL) is gratefully acknowledged.
\end{acknowledgments}

\appendix

\section{Effects of higher orders in the strains}
\label{app:nonlinear}

A priori, a formulation of stretching energy densities in terms of
the variables $\eta$ and $\eta_z$ provides a more adequate
framework for discussing the incompressible limit than a
formulation in terms of the respective linearized expressions
$u_{ii}$ and $u_{zz}$. Using $u_{ii}$ and $u_{zz}$, on the other
hand, makes our models more tractable and perhaps also more
intuitive because we stay in close contact to the standard
formulation of elasticity as presented in textbooks. The purpose
of this appendix is to discuss what changes occur in our theories
if we use $\eta$ and $\eta_z$ instead of $u_{ii}$ and $u_{zz}$.

\subsection{Biaxial elastomers}

Now we use Eq.~(\ref{uniEnNonlinear}) as the starting point for
setting up our model for soft biaxial elastomers. When $C_4$ can
become negative, higher order terms featuring $\hat{u}_{ab}$,
$\eta$ and $\eta_z$ must be added to Eq.~(\ref{uniEnNonlinear}) so
that the model elastic energy density becomes
\begin{align}
\label{uniEn1Nonlinear} f_{\text{uni}}^{(2)} = f_{\text{uni}} +
A_1  \, \eta_z \, \hat{u}_{ab}^2 + A_2 \,  \eta \, \hat{u}_{ab}^2
+ B \, (\hat{u}_{ab}^2)^2 \, .
\end{align}
Proceeding in close analogy to the steps described following
Eq.~(\ref{uniEn1}), we find that the equilibrium values of $\eta$
and $\eta_z$ are
\begin{subequations}
\label{equiVals}
\begin{align}
\label{equiEtaZ} \eta_z^0 &=  \alpha \, ( \hat{u}_{ab}^0)^2 =
\textstyle{\frac{1}{2}}\, \alpha \, S^2 \, ,
\\
\label{equiEta} \eta^0  &=  \beta \, ( \hat{u}_{ab}^0)^2 =
\textstyle{\frac{1}{2}}\, \beta \, S^2 \, ,
\end{align}
\end{subequations}
with $\alpha$ and $\beta$ as given in
Eq.~Eq.~(\ref{resAlphaBeta}). The equilibrium values $u^0_{az}$
and $\hat{u}_{ab}^0$ remain unchanged. To learn more about the
equilibrium state, our next task is to determine the equilibrium
strain tensor $\tens{u}^0$. Using our knowledge about $u^0_{az}$
and $\hat{u}_{ab}^0$, it is clear that $\tens{u}^0$ is of the form
\begin{align}
\tens{u}^0 = \left(
\begin{array}{ccc}
\frac{1}{2} (t+S) &0 & 0\\
0 & \frac{1}{2} (t-S) &0\\
0 & 0 & u_{zz}^0
\end{array}
\right) ,
\end{align}
where we used the abbreviation $t = u_{cc}^0$. $t$ and $u_{zz}^0$
are unknown thus far and we need to determine them as functions of
the order parameter $S$. Replacing $\eta_z$ and $u_{zz}$ in
Eq.~(\ref{defEtaZ}) by $\eta_z^0$ and $u_{zz}^0$, using
Eq.~(\ref{equiEtaZ}) for $\eta_z^0$ and solving for $u_{zz}^0$ we
find
\begin{align}
u_{zz}^0 =  \frac{1}{2}\, \left[ \left(  1 +
\textstyle{\frac{1}{2}}\, \alpha \, S^2 \right)^2 -1  \right] .
\end{align}
To leading order in $S$ this result reduces to $u_{zz}^0 =
\textstyle{\frac{1}{2}}\, \alpha \, S^2$ which coincides with the
result of our linearized theory presented in
Sec.~\ref{biaxialElastomers}. It remains to calculate $t$. Since
$\tens{u}^0$ is diagonal, Eq.~(\ref{defEta}) leads simply to
\begin{align}
\eta^0 = \left[ (1+t+S) (1+t-S)(1 + 2 u_{zz}^0) \right]^{1/2} -1
\, .
\end{align}
Using Eqs.~(\ref{equiVals}) for $\eta^0$ and $u_{zz}^0$ we obtain
by solving for $t$:
\begin{align}
\label{resT} t =  \left[ \frac{(1 + \frac{\beta}{2} \, S^2)^2}{(1
+ \frac{\alpha}{2} \, S^2)^2} + S^2 \right]^{1/2} -1 \, .
\end{align}
To leading order in $S$ this expression reduces to $t =
\frac{1}{2} (\beta - \alpha +1) S^2$. Comparing this to the result
$u_{cc}^0 =  \frac{1}{2} (\beta - \alpha ) S^2$ of our linearized
theory, we see that the linearized theory lacks the contribution
$\frac{1}{2} S^2$ to $u_{cc}^0$. Nevertheless, up to this detail,
the form of the equilibrium state predicted by the linearized and
the non-linearized theory is the same. In the end, the
non-linearized theory leads to the elastic energy density
\begin{align}
f = & \textstyle{\frac{1}{2}} C_1 g_{zz}^{-2} ( \delta u_{zz})^2 +
C_2 g_{zz}^{-1} \overline{u} \delta u_{zz} +
\textstyle{\frac{1}{2}}C_3 (\overline{u})^2 \nonumber \\
& + C_5 ( \delta u_{az})^2 + B_2 S^2 (\delta u_{xx} - \delta
u_{yy})^2 & \nonumber \\
& + (A_1 S g_{zz}^{-1/2} \delta u_{zz} + A_2 S \overline{u} )
(\delta u_{xx} - \delta u_{yy} ) , \label{f_biax_nonlinear}
\end{align}
for the soft biaxial state, where
\begin{equation}
\overline{u} = [\det (1 + 2 \tens{u}_0 )]^{1/2} \,\,
[(\tens{\delta} + 2 \tens{u}_0)^{-1} \delta \tens{u} ]_{ii} =
\frac{V_B}{V_0} u'_{ii} \approx \frac{\delta V}{V_0} ,
\end{equation}
where $V_0$ is the volume of the reference uniaxial state, $V_B$
is the volume of the biaxial state and $\delta V = V - V_B$. Thus,
volume changes in the biaxial phase are suppressed by the term
proportional to $C_3$ in Eq.~(\ref{f_biax_nonlinear}).  In the
incompressible limit $C_3 \rightarrow \infty$, the nonlinear
theory indeed produces fixed $\delta V = 0$ and not $\delta u_{ii}
= 0$ as the linearized theory does. Our findings about the
softness of the biaxial state, however, remain practically the
same.

\subsection{Sm$C$-elastomers}

When $C_5$ can become negative in response to Sm$C$ ordering of
the mesogens, higher order terms featuring $u_{az}$, $\eta$ and
$\eta_z$ must be added to Eq.~(\ref{uniEnNonlinear}) to ensure
mechanical stability. Then the model elastic energy density
becomes
\begin{align}
\label{uniEn2Nonlinear} f_{\text{uni}}^{(2)} &= f_{\text{uni}} +
D_1\,\eta_z\, u_{az}^2 + D_2 \,  \eta \, u_{az}^2 + D_3 \,
\hat{u}_{ab} u_{az} u_{bz}
\nonumber \\
&+ E  \, (u_{az}^2)^2 \, .
\end{align}
Essentially repeating the algebra described following
Eq.~(\ref{uniEn2}), we find
\begin{subequations}
\label{equiVals2}
\begin{align}
\label{equiEtaZ2} \eta_z^0 &=  \sigma \, (u_{az}^0)^2 = \sigma \,
S^2 \, ,
\\
\label{equiEta2} \eta^0  &=  \tau \, ( u_{az}^0)^2 = \tau \, S^2
\, ,
\end{align}
\end{subequations}
with $\sigma$ and $\tau$ as given in  Eq.~(\ref{resSigmaTau}).
$\hat{u}_{ab}^0$ and $u_{az}^0$ remain the same as in
Sec.~\ref{smecticElastomers1Transition}. Consequently, the
equilibrium strain tensor is of the form
\begin{align}
\label{formU0SmC} \tens{u}^0 = \left(
\begin{array}{ccc}
\frac{t}{2} &0 & S\\
0 & \frac{t}{2} &0\\
S & 0 & u_{zz}^0
\end{array}
\right) ,
\end{align}
where $t = u_{cc}^0$. For determining the equilibrium state, it
remains to compute $t$ and $u_{zz}^0$ as functions of $S$. From
Eqs.~(\ref{defEtaZ}) and (\ref{equiEtaZ2}) it follows readily that
\begin{align}
u_{zz}^0 =  \frac{1}{2}\, \left[ \left(  1 + \sigma \, S^2
\right)^2 -1  \right] ,
\end{align}
which is to leading order in $S$ identical to the result $u_{zz}^0
= \sigma \, S^2$ of our linearized theories for Sm$C$ elastomers.
Equations~(\ref{defEtaZ}) and (\ref{formU0SmC}) lead us to
\begin{align}
\eta^0 = \left[ (1+ t)^2 (1 + 2 \, u_{zz}^0) - 4 \, (1+ t)
S^2\right]^{1/2} -1 \, .
\end{align}
Taking into account the equilibrium values~(\ref{equiEtaZ2}) and
solving for $t$ we find
\begin{align}
t =  \frac{2 S^2}{(1+\sigma S^2)^2} +  \left[  \frac{(1+\tau
S^2)^2}{(1+\sigma S^2)^2}  + \frac{4 S^4}{(1+\sigma S^2)^4}
\right]^{1/2} -1 \,  .
\end{align}
To leading order in $S$, $t = (\tau - \sigma +2)\, S^2$ which has
to be compared to the $u_{cc}^0 =  (\tau - \sigma )\, S^2$
stemming from our linearized theories for Sm$C$ elastomers. Note
that the linearized theories miss the contribution $2S^2$ to
$u_{cc}^0$. However, as was the case for the biaxial soft state,
this discrepancy does not affect the nature of the softness.

\section{Elastic constants}
\label{app:elasticConstants} This appendix collects our results
for the elastic constants of biaxial and Sm$C$ elastomers.

\subsection{Elastic constants of biaxial elastomers}
\label{app:elasticConstantsBiaxial}

The elastic constants of soft biaxial elastomers as defined in the
elastic energy density~(\ref{biaxEn}) are given to order $O(S^3)$
by
\begin{align}
C_{zzzz}  &=  C_1 + 2C_2 + C_3 + 2 \alpha [C_1 + 2C_2 + C_3] \,
S^2  \, ,
\\
C_{xzxz} &=  2[  C_5 + C_5 \, S + \textstyle{\frac{1}{2}} (\alpha
+ \beta) C_5 \, S^2 ] \, ,
\\
C_{yzyz} & = 2[ C_5 - C_5 \, S + \textstyle{\frac{1}{2}} (\alpha +
\beta) C_5 \, S^2 ]\, ,
 \\
 C_{zzxx}& =  C_2 + C_3 + [C_2 + C_3 + A_1 + A_2]\, S
 \nonumber \\
 &+ [\textstyle{\frac{1}{2}} (\alpha + \beta) (C_2 + C_3) + A_1 + A_2] \, S^2  \, ,
\\
C_{zzyy} &= C_2 + C_3 - [C_2 + C_3 + A_1 + A_2]\, S
 \nonumber \\
 &+ [\textstyle{\frac{1}{2}} (\alpha + \beta) (C_2 + C_3) + A_1 + A_2] \, S^2 \, ,
\\
C_{xxxx} &= C_3 + 2 [ C_3 + A_2 ] \, S
\nonumber\\
&+ [(1-\alpha +\beta)C_3 + 4 A_2 + 2 B] \, S^2\, ,
\\
C_{yyyy} &=  C_3 - 2 [ C_3 + A_2 ] \, S
\nonumber\\
&+ [(1-\alpha +\beta)C_3 + 4 A_2 + 2 B] \, S^2 \, ,
\\
C_{xxyy} &= C_3 - [(1+\alpha -\beta) C_3 + 2 B]\, S^2\, .
\end{align}

\subsection{Elastic constants of Sm$C$ elastomers}
\label{app:elasticConstantsSmC} Here we list our results for the
elastic constants of soft Sm$C$ elastomers as defined in
Eq.~(\ref{C2hEn}).

\subsubsection{Elastic constants as obtained from the strain-only
model of Sec.~\ref{smecticElastomers1}}

Our results for the angle $\theta$ and the elastic constant
$\bar{C}$ read
\begin{align}
\theta &= \tan^{-1} \left( \frac{- \Lambda_{zz}^0 \omega \,
S}{\Lambda_{xx}^0 - \Lambda_{xz}^0 \omega \, S} \right) = - \omega
\, S + O(S^3)  \, ,
\\
\bar{C} &= 4 C_4 \, \frac{(\Lambda_{yy}^0)^2 (\Lambda_{xx}^0 -
\Lambda_{xz}^0 \omega \, S)^2}{\cos^2 \theta}
\nonumber \\
& = 4 C_4  + 4 C_4 [2(\tau-\sigma) + (\omega -4)\omega] \, S^2 +
O(S^4)\, .
\end{align}
For the remaining elastic constants we find to order $O(S^3)$
\begin{align}
C_{zzzz} &= C_1 + 2 C_2 + C_3 + 4 (\sigma -2 ) (C_1 + 2 C_2 +
C_3)\,  S^2\,  ,
\\
C_{xzxz} &= 8 [2 (C_1 + 2 C_2 + C_3) + 2(D_1 + D_2) + E] \, S^2\,
,
\\
C_{zzxx} &= C_2 + C_3 + [ (\sigma + \tau +\omega -4) (C_2 + C_3)
\nonumber\\
&+ 4 (C_1 + 2 C_2 + C_3) + 4 (D_1 + D_2) ]\, S^2\, ,
\\
C_{zzyy} &=  C_2 + C_3 + (\sigma +\tau -\omega -4) (C_2 + C_3) \,
S^2  \, ,
\\
C_{xxxx} &= C_3 + C_4 + 2 [4 C_2 + (4+\tau +\omega -\sigma)C_3
\nonumber\\
&+ (\tau +\omega -\sigma)C_4 + 2(2D_2 + D_3)]\, S^2 \, ,
\\
C_{yyyy} &= C_3 + C_4 + 2 (\tau-\omega-\sigma)(C_3 + C_4) \, S^2
\, ,
\\
C_{xxyy} &= C_3 - C_4 + 2[ 2 (C_2 + C_3) -  (\sigma -\tau)(C_3 -
C_4)
\nonumber\\
&+ 2D_2 -D_3 ]\, S^2 \, ,
\\
C_{xxxz} &= [4 (C_2 + C_3) + 2D_2 +D_3]\, S \, ,
\\
C_{yyxz} &= [4 (C_2 + C_3) + 2D_2 -D_3]\, S  \, ,
\\
C_{zzxz} &= 2 [ 2 (C_1 + 2 C_2 + C_3) +D_1 + D_2] \, S \,  .
\end{align}

\subsubsection{Elastic constants as obtained from the model with
strain and director of Sec.~\ref{smecticElastomers2}}

For $\theta$ and $\bar{C}$ we find
\begin{align}
\theta &= \tan^{-1} \left( \frac{- \Lambda_{zz}^0 \,
\Pi}{\Lambda_{xx}^0 \, \Xi - \Lambda_{xz}^0 \, \Pi} \right) = -
\frac{\bar{\omega}}{\bar{\rho}} \, S + O(S^3)  \, ,
\\
\bar{C} &= 4 \, \frac{C_4 C_5}{\Delta} \, \frac{(\Lambda_{yy}^0)^2
\, (\Lambda_{xx}^0 \, \Xi  - \Lambda_{xz}^0 \, \Pi)^2}{\cos^2
\theta}
\nonumber \\
& = 4 C_4  + 4 \frac{2 \bar{\omega}^2 C_4 + [2 \bar{\rho}^2
(\bar{\sigma} - \bar{\tau} + 2 \bar{\omega}) - \bar{\omega}^2]
C_5}{\bar{\rho}^2 \, C_5} \, C_4\, S^2
\nonumber \\
&+ O(S^4)\, .
\end{align}
The remaining elastic constants are given to order $O(S^3)$ by
\begin{align}
C_{zzzz} &= C_1 + 2 C_2 + C_3 + 4 \bigg[(\bar{\sigma} -2
\bar{\rho}^2)
\nonumber\\
&\times (C_1 + 2 C_2 + C_3) - 2 \frac{(\lambda_1 +
\lambda_2)^2}{\lambda_4^2}C_5\bigg] \,  S^2\,  ,
\\
C_{xzxz} &= 8 \bigg[2 \bar{\rho}^2 (C_1 + 2 C_2 + C_3)
\nonumber\\
& + \frac{g C_5 + \lambda_4[\lambda_4 - 2 (2 \lambda_2 -
\lambda_3) ]}{\lambda_4^2}C_5 \bigg] \, S^2\, ,
\\
C_{zzxx} &= C_2 + C_3 + \bigg[ 4 \bar{\rho}^2 C_1+ (4 \bar{\rho}^2
+\bar{\sigma} + \bar{\tau} +\bar{\omega}) C_2
\nonumber \\
&+ (\bar{\sigma} + \bar{\tau} +\bar{\omega})C_3 - 4 \frac{2
\lambda_2^2 + \lambda_1 (2\lambda_2 + \lambda_3)}{\lambda_4^2} C_5
\nonumber\\
&+ 4 \frac{\bar{\rho} \lambda_3 \lambda_4 - \lambda_2 (\lambda_3 +
2 \bar{\rho} \lambda_4)}{\lambda_4^2} C_5 \bigg]\, S^2 ,
\\
C_{zzyy} &=  C_2 + C_3 + \bigg[ (\bar{\sigma} +\bar{\tau}
-\bar{\omega} -4\bar{\rho}^2) (C_2 + C_3)
\nonumber \\
&- 4 \frac{(\lambda_1 + \lambda_2)(2 \lambda_2 -
\lambda_3)}{\lambda_4^2}C_5 \bigg] \, S^2  \, ,
\\
C_{xxxx} &= C_3 + C_4 + 2 \bigg[  4 \bar{\rho}^2(C_2 + C_3)
\nonumber \\
&+(\bar{\tau} +\bar{\omega} -\bar{\sigma}) (C_3 + C_4)
\nonumber\\
&-  \frac{(2\lambda_2 + \lambda_3)^2 + 8 \bar{\rho}
\lambda_4(\lambda_1 + \lambda_2)}{\lambda_4^2}  C_5 \bigg]   S^2\,
,
\\
C_{yyyy} &= C_3 + C_4 + 2 \bigg[
(\bar{\tau}-\bar{\omega}-\bar{\sigma})(C_3 + C_4)
\nonumber \\
&+ \frac{(2 \lambda_2 - \lambda_3)^2}{\lambda_4^2} C_5 \bigg]  \,
S^2 \, ,
\\
C_{xxyy} &= C_3 - C_4 + 2\bigg[ 2 \bar{\rho}^2 (C_2 + C_3) -
(\bar{\sigma} -\bar{\tau})(C_3 - C_4)
\nonumber\\
&-2 \frac{(\lambda_1 + \lambda_2)(2 \lambda_2 - \lambda_3) +
\bar{\rho} \lambda_4 (2 \lambda_2 + \lambda_3)}{\lambda_4^2} C_5
\bigg]\, S^2 \, ,
\\
C_{xxxz} &= 4\bigg[\bar{\rho} (C_2 + C_3) - \frac{\lambda_1 +
\lambda_2}{\lambda_4} C_5\bigg]\, S \, ,
\\
C_{yyxz} &= 2 \bigg[2 \bar{\rho} (C_2 + C_3) - \frac{2 \lambda_2 +
\lambda_3}{\lambda_4} C_5\bigg]\, S  \, ,
\\
C_{zzxz} &= 2 \bigg[ 2\bar{\rho} (C_1 + 2 C_2 + C_3) - \frac{2
\lambda_2 - \lambda_3}{\lambda_4} C_5 \bigg] \, S \,  .
\end{align}

\section{Steps leading to Eq.~(\ref{resBeforeRelaxation})}
\label{app:steps} In this appendix we outline some of the
algebraic steps leading from Eqs.~(\ref{f1expanded}) and
(\ref{f2expanded}) to Eq.~(\ref{resBeforeRelaxation}). As
discussed in the text following Eq.~(\ref{f2}), the role of
$\delta \tilde{c}_y$ is special in that it is the local relaxation
of this quantity that makes Sm$C$ elastomers soft. To see this, we
recast Eq.~(\ref{f1expanded}), where $\delta \tilde{c}_y$ appears
in two different terms, by using
\begin{align}
\label{algebraIdentity} &2 C_4  \left( \delta \bar{w}_{xy}
\right)^2 + C_5 \left( \delta \bar{w}_{yz} \right)^2
\nonumber\\
&= \Delta \left[  \delta \tilde{c}_y +\frac{  2 C_4 \, \Pi \,
\delta u_{xy} + C_5 \, \Xi  \, \delta u_{yz} }{\Delta} \right]^2
\nonumber\\
&+ 2\, \frac{C_4 C_5}{\Delta} \, \left[  \Pi \,  \delta u_{yz} -
\Xi  \, \delta u_{xy} \right]^2 ,
\end{align}
with $\Pi$, $\Xi$, and $\Delta$ as defined below
Eq.~(\ref{resBeforeRelaxation}). The validity of
Eq.~(\ref{algebraIdentity}) can be checked by straightforward but
slightly tedious algebra.

The second major step in going from Eqs.~(\ref{f1expanded}) and
(\ref{f2expanded}), after these are combined, to
Eq.~(\ref{resBeforeRelaxation}) is to integrate out the massive
variable $\delta \tilde{c}_x$, i.e., to replace  $\delta
\tilde{c}_x$ by its minimum value. This minimization is
straightforward but a little inconvenient because of the number of
terms that are involved. Combining Eq.~(\ref{algebraIdentity}) and
the outcome of this minimization we arrive at the intermediate
result
\begin{widetext}
\begin{align}
\delta  f &= \Delta \left[  \delta \tilde{c}_y +\frac{  2 C_4 \,
\Pi \, \delta u_{xy} + C_5 \, \Xi  \, \delta u_{yz} }{\Delta}
\right]^2 + 2\, \frac{C_4 C_5}{\Delta} \, \left[  \Pi \, \delta
u_{yz} - \Xi  \, \delta u_{xy} \right]^2
\nonumber \\
& + \frac{1}{2}  \bigg[C_1 + 2C_2 + C_3 - \frac{2}{\Omega}
(\lambda_1 + \lambda_2)^2 S^2\bigg]  (\delta u_{zz})^2 +  \bigg[
C_5 - \frac{\lambda_4^2}{4\Omega} \left(  1 - \frac{3}{2}
S^2\right) \bigg]  (\delta u_{xz})^2
\nonumber \\
& +  \frac{1}{2}\bigg[C_3 +C_4 - \frac{2}{\Omega} \left(\lambda_2
+\frac{\lambda_3}{2} \right)^2 S^2 \bigg]   (\delta u_{xx})^2
+\frac{1}{2} \bigg[C_3 +C_4 +\frac{2}{\Omega} \left(\lambda_2
-\frac{\lambda_3}{2} \right)^2 S^2 \bigg] (\delta u_{yy})^2
\nonumber \\
&+ \bigg[C_2 + C_3 - \frac{2}{\Omega} (\lambda_1 + \lambda_2)
\left(\lambda_2 +\frac{\lambda_3}{2} \right) S^2  \bigg] \delta
u_{zz}  \delta u_{xx} +  \bigg[C_2 + C_3 - \frac{2}{\Omega}
(\lambda_1 + \lambda_2)   \left(\lambda_2 -\frac{\lambda_3}{2}
\right) S^2  \bigg] \delta u_{zz} \delta u_{yy}
\nonumber \\
& + \bigg[C_3 - C_4  - \frac{2}{\Omega} (\lambda_1 + \lambda_2)
\left(\lambda_2 -\frac{\lambda_3}{2} \right) S^2 \bigg] \delta
u_{xx} \delta u_{yy} + \bigg[ - \frac{1}{\Omega} (\lambda_1 +
\lambda_2) \lambda_4 \,S \left(  1 - \frac{3}{2} S^2\right) \bigg]
\delta u_{xx} \delta u_{xz}
\nonumber \\
& +  \bigg[ - \frac{1}{\Omega} \left(\lambda_2
+\frac{\lambda_3}{2} \right) \lambda_4 \,S \left(  1 - \frac{3}{2}
S^2\right) \bigg]  \delta u_{yy} \delta u_{xz} +  \bigg[ -
\frac{1}{\Omega} \left(\lambda_2 - \frac{\lambda_3}{2} \right)
\lambda_4 \,S \left(  1 - \frac{3}{2} S^2\right) \bigg]  \delta
u_{zz} \delta u_{xz} \, ,
\end{align}
where we have used the shorthand $\Omega = g S^2 + C_5
\bar{\rho}^2 (1+ 5 S^2/2)$. Switching from the strain variable $\delta \tens{u}$ to $\tens{u}'=
(\tens{\Lambda}^{0T})^{-1} \delta \tens{u} \, (\tens{\Lambda}^{0})^{-1}$ then takes us to Eq.~(\ref{resBeforeRelaxation}).
\end{widetext}

\section{Effects of the Frank energy on the softness}
\label{app:frank} In Sec.~\ref{smecticElastomers2} we have
neglected the Frank energy. In this appendix we check if our
conclusions about the softness of Sm$C$ elastomers are affected if
we take the Frank energy, or rather the corresponding density
\begin{align}
\label{FrankEnergy} f_{\text{Frank}} &= \textstyle{\frac{1}{2}}
K_1 \, [\nabla \cdot \brm{n}]^2 + \textstyle{\frac{1}{2}} K_2 \, [
\brm{n} \cdot (\nabla \times \brm{n})]^2
\nonumber \\
&+ \textstyle{\frac{1}{2}} K_3 \, [ \brm{n} \times (\nabla \times
\brm{n})]^2 ,
\end{align}
into account. To this end we have to check if
Eq.~(\ref{FrankEnergy}) can lead to a mass for $\delta
\tilde{c}_y$.

Each of the three terms in Eq.~(\ref{FrankEnergy}) features a
derivative. Since the director $\brm{n}$ lives in the target
space, these derivatives are derivatives with respect to the
target space coordinate $\brm{R}$, i.e.,
\begin{align}
\nabla_k n_i = \frac{\partial n_i}{ \partial R_k} =
\big(\tens{\Lambda}^{-1} \big)_{lk} \frac{\partial n_i}{ \partial
x_l} \, .
\end{align}
In Sec.~\ref{smecticElastomers2} we modeled Sm$C$ elastomers via
converting the director to a reference space vector by using the
polar decomposition theorem. Doing the same conversion here we get
with the help of Eq.~(\ref{targetToRefernce})
\begin{align}
\nabla_k n_i = \big(\tens{\Lambda}^{-1} \big)_{lk} \left\{
\frac{\partial O_{ij}}{ \partial x_l} \, \tilde{n}_j + O_{ij} \,
\frac{\partial \tilde{n}_j}{ \partial x_l} \right\} \, .
\end{align}
The second term in the braces contains a derivative of
$\tilde{n}_j$ and hence it is clear that it cannot lead to a
massive term. This leaves us with the first term in the braces as
a potential source of a mass. Using Eq.~(\ref{ExpRotMatrix}) we
get
\begin{align}
\frac{\partial O_{ij}}{ \partial x_l} =  \frac{\partial }{
\partial x_l} \, \eta_{Aij} + \cdots \, ,
\end{align}
where $\eta_{Aij} = \frac{1}{2} (\partial_j u_i - \partial_i u_j)$
is the antisymmetric part of the displacement  gradient tensor
$\eta_{ij} = \partial_j u_i$. Thus, the first term in the braces couples $\tilde{n}_j$ to derivatives of $\eta_{ij}$ and higher order terms, and as such this term cannot make $\delta \tilde{c}_y$ massive in a perturbative (diagrammatic) expansion.
Furthermore, the terms in this expansion will be subdominant
compared to the $u_{yz} {\tilde c}_y$ at small wavenumber $q$.
Rotational invariance in the reference space should dictate that
Frank energy terms do not ever generate a mass for ${\tilde c}_y$,
a full demonstration of this fact to all orders in perturbation
theory is beyond the scope of this paper. We conclude that our findings about the soft elasticity of Sm$C$ elastomers remain unchanged if we include the Frank energy in our
model.


\begin{thebibliography}{}
\bibitem{stress_tensors}
The stress tensor $\sigma_{xy}$ defined in Eq.\ (\ref{stress_1})
is the second Piola-Kirchhoff stress tensor.  It transforms like a
tensor in the reference space but like a scalar under rotations in
the target space.  The more familiar Cauchy stress tensor, which
transforms like a tensor in the target space is $\sigma_{ij}^C =
(\det \tens{\Lambda})^{-1} \Lambda_{ik} \sigma_{kl}
\Lambda^T_{lj}$. The engineering or first Piola-Kirchhoff stress
tensor, $\sigma_{ij}^{\text{eng}} = \Lambda_{ik} \sigma_{kj}$, is
a mixed tensor that transform like a vector under rotations in
either the reference or the targt space.
\bibitem{footnoteSmallRotation}
In order to ensure the full invariance under $\tens{O}_R$ we would
have to include higher order terms.
\bibitem{fourierTransformation}
We define spatial Fourier transforms via $A (\brm{x}) =
(2\pi)^{-3} \int d^3 q \, \widetilde{A} (\brm{q}) \, \exp (i
\brm{q} \cdot \brm{x})$.
\end{thebibliography}
\end{document}